\documentclass[journal]{IEEEtran}

\makeatletter
\def\ps@headings{%
	\def\@oddhead{\mbox{}\scriptsize\rightmark \hfil \thepage}%
	\def\@evenhead{\scriptsize\thepage \hfil \leftmark\mbox{}}%
	\def\@oddfoot{}%
	\def\@evenfoot{}}
\makeatother 

\IEEEoverridecommandlockouts

\usepackage{amsfonts}
\usepackage[dvips]{graphicx}
\usepackage{amsmath}
\usepackage[caption=false,font=normalsize,labelfont=sf,textfont=sf]{subfig}
\usepackage{times}
\usepackage{cite}
\usepackage{lettrine}
\allowdisplaybreaks[4]
\usepackage{array}
\usepackage{amssymb}

\usepackage{stfloats}
\usepackage{graphicx}
\usepackage{footnote}
\usepackage{booktabs}
\usepackage{array}
\usepackage{algorithmic}
\usepackage{algorithm}
\usepackage{subeqnarray}
\usepackage{cases}
\usepackage{threeparttable}
\usepackage{color}
\usepackage{bm}
\usepackage{bbm}

\usepackage{capt-of}    
\usepackage{tabularx}
\usepackage{multirow}

\newtheorem{theorem}{\underline{Theorem}}[section]
\newtheorem{lemma}{\underline{Lemma}}[section]

\newtheorem{proposition}{Proposition}[section]

\newtheorem{remark}{\underline{Remark}}[section]

\begin{document}
	\bibliographystyle{IEEEtran}
	
	\title{
		Relay-Assisted Cooperative Federated Learning
	}
	
	\IEEEoverridecommandlockouts
	\author{
		Zehong Lin,~\IEEEmembership{Graduate Student Member,~IEEE}, Hang Liu,~\IEEEmembership{Member,~IEEE}, \\and Ying-Jun Angela Zhang,~\IEEEmembership{Fellow,~IEEE}
		
		\thanks{This article was presented in part at the IEEE Global Communications Conference (GLOBECOM), Madrid, Spain, in December 2021 \cite{GC2021}.

        The authors are with the Department of Information Engineering, The Chinese University of Hong Kong, Hong Kong (e-mail: lz018@ie.cuhk.edu.hk; lh117@ie.cuhk.edu.hk; yjzhang@ie.cuhk.edu.hk)}
	}
	\maketitle
	
	\begin{abstract}
		\emph{Federated learning} (FL) has recently emerged as a promising technology to enable artificial intelligence (AI) at the network edge, where distributed mobile devices collaboratively train a shared AI model under the coordination of an edge server. To significantly improve the communication efficiency of FL, \emph{over-the-air computation} allows a large number of mobile devices to concurrently upload their local models by exploiting the superposition property of wireless multi-access channels. Due to wireless channel fading, the model aggregation error at the edge server is dominated by the weakest channel among all devices, causing severe straggler issues. In this paper, we propose a \emph{relay-assisted cooperative FL} scheme to effectively address the straggler issue. In particular, we deploy multiple half-duplex relays to cooperatively assist the devices in uploading the local model updates to the edge server. The nature of the over-the-air computation poses system objectives and constraints that are distinct from those in traditional relay communication systems. Moreover, the strong coupling between the design variables renders the optimization of such a system challenging. To tackle the issue, we propose an alternating-optimization-based algorithm to optimize the transceiver and relay operation with low complexity. Then, we analyze the model aggregation error in a single-relay case and show that our relay-assisted scheme achieves a smaller error than the one without relays provided that the relay transmit power and the relay channel gains are sufficiently large. The analysis provides critical insights on relay deployment in the implementation of cooperative FL. Extensive numerical results show that our design achieves faster convergence compared with state-of-the-art schemes.
	\end{abstract}
	
	\section{Introduction}
	\IEEEPARstart{W}{ith} the proliferation of mobile devices, such as smartphones, wearable devices, and sensors, mobile data traffic is growing at an unprecedented rate. The massive amounts of mobile data stimulate the development of artificial intelligence (AI) technologies at the edge of wireless networks and trigger the emergence of a new paradigm termed \emph{edge intelligence} (EI) \cite{zhou2019edge, letaief2019roadmap, park2019wireless, shi2020communication}, including edge learning and edge inference. With the support of mobile edge computing \cite{mao2017survey}, EI pushes the training and inference processes of AI models from central cloud servers to the network edge, i.e., edge servers and mobile devices. Due to the close proximity and fast access to data sources, EI significantly reduces network bandwidth/energy costs and improves privacy preservation.
	
	Federated learning (FL) \cite{konevcny2016federated2} is a promising framework to achieve edge learning by exploiting the increasingly powerful computational capabilities of mobile devices. Under the coordination of an edge server, e.g., a base station (BS) or a wireless access point (AP), FL enables mobile devices to collaboratively train a shared model without transmitting the raw data to the edge server. Specifically, the edge server shares a global model with mobile devices for distributed on-device training using their local data. Then, the mobile devices upload their local model updates to the edge server to update the global model as a weighted sum of the local models. Compared with traditional centralized learning approaches that collect all data at a central server, FL avoids prohibitive data transmission delay and potential privacy disclosure. Nonetheless, the communication cost remains the main bottleneck in FL since the learning process critically relies on iterative model exchanges between the edge server and the distributed mobile devices over limited radio resources.
	
	To relieve the communication bottleneck in FL, over-the-air computation \cite{nazer2007computation} has been introduced to support concurrent model uploading over the same radio resources by exploiting the signal-superposition property of a wireless multiple-access channel. With proper pre-equalization and transmit power control at the mobile devices, the local models received at the edge server are aligned as a linear combination that combats the wireless fading and achieves the desired weighted sum \cite{zhu2018mimo, liu2020over, cao2020optimized, zhang2021gradient}. Unlike conventional orthogonal multiple access schemes that allocate orthogonal channels to multiple mobile devices for independent transmissions, the required bandwidth or communication latency of over-the-air computation is independent of the number of devices, significantly enhancing the system scalability and improving the communication efficiency. Despite the above advantages, over-the-air model aggregation inevitably suffers from distortion caused by communication noise and wireless channel fading. To align the uploaded local model updates at the edge server, the devices with better channels have to lower the transmit power, making the aggregation error dominated by the devices with weak channels, i.e., the communication stragglers. Existing work \cite{zhu2019broadband, zhu2020one, yang2020federated, amiri2020federated} avoided large model aggregation errors by excluding the stragglers from concurrent model uploading. For instance, the authors in \cite{zhu2019broadband} proposed a truncated-based power control approach to discard devices with deep fading channels. In \cite{yang2020federated}, the authors considered joint device selection and beamforming design to maximize the number of selected devices under a target mean-square-error (MSE) requirement. However, discarding devices from training often decreases the number of exploited training data, which inevitably slows down the FL convergence \cite{liu2020reconfigurable}.
		
	To overcome the straggler issue and avoid potential learning performance loss, an alternative approach is not to exclude the communication stragglers but to enhance their communication qualities by advanced communication technologies. For example, relay-assisted communication has been a cost-effective technique to combat the effects of channel fading in wireless communications systems \cite{sendonaris2003user1, sendonaris2003user2, laneman2004cooperative}. Specifically, we can deploy a set of relays as intermediate nodes to cooperatively assist data exchanges between the mobile devices and the AP, i.e., the edge server, which extends the coverage of the AP and enhances the transmission reliability of the devices far away from the AP. Inspired by this, the authors in \cite{feng2019joint} built a relay network to improve the energy efficiency in FL, where the mobile devices cooperatively form a relay network to help each other upload the model updates to the AP. Recent work in \cite{qu2021partial} proposed a two-tier relay-assisted FL framework and designed a partially synchronized parallel mechanism, where the models and local gradients are transmitted simultaneously and aggregated at relay nodes. Note that Refs. \cite{feng2019joint, qu2021partial} assume that each device only connects to one relay node and cannot leverage the benefits of the \emph{cooperative diversity} of relay networks \cite{laneman2004cooperative}. The authors in \cite{wang2020optimized} exploited the cooperative diversity with multiple relays for over-the-air computation and proposed a hierarchical communication framework to minimize the communication MSE. Nonetheless, Ref. \cite{wang2020optimized} considered a general over-the-air computation task and did not examine the effectiveness of the design in FL systems. More importantly, similarly to the relay-assisted design in conventional communication systems, the above work \cite{feng2019joint, qu2021partial, wang2020optimized} assumes that the devices do not transmit to the edge server directly. Instead, the local model uploading in each communication round is divided into two phases. The devices transmit their model information to the relays in the first phase, and rely on the relays to forward the information to the server in the second phase. This restriction, however, leads to sub-optimal learning performance in FL systems, as the AP fails to exploit the information carried in the direct transmission by the devices. Notice that the signals received at the AP from multiple transmitters can be coherently aligned as a weighted sum for over-the-air model aggregation. Therefore, with proper configuration of device-side and relay-side pre-equalization, the AP can aggregate the signals received from the mobile devices and the relays across the two time slots to achieve cooperative diversity gain. Such cooperation of devices and relays can significantly enhance the performance of over-the-air model aggregation, especially when the relay-only design is insufficient. Notably, the design of the device-relay cooperative system requires joint optimization of the equalization factors at the devices, the relays, and the server.

	\begin{figure}[t]
		\centering
		\includegraphics[width=0.95\linewidth]{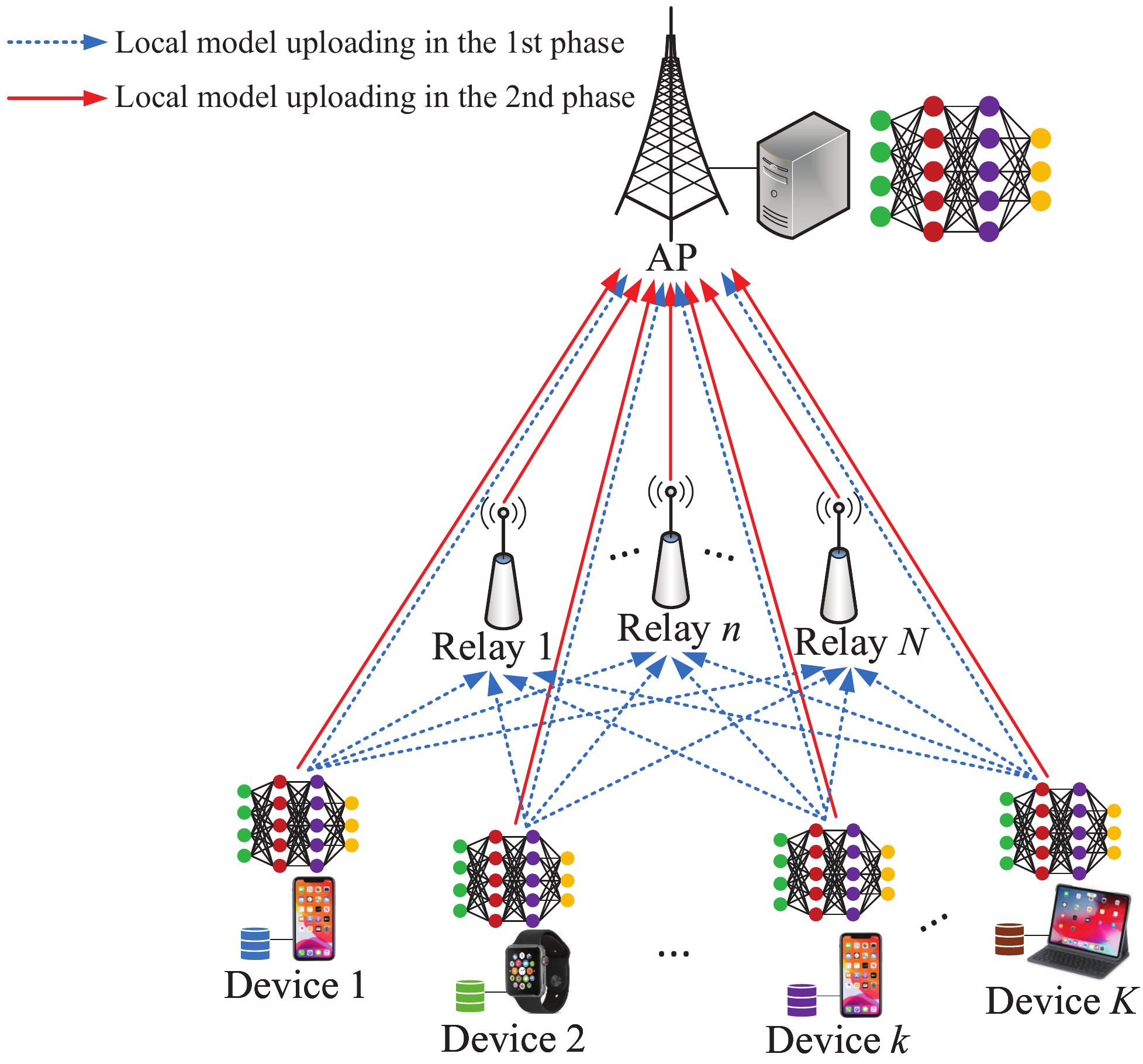}
		\caption{The considered relay-assisted cooperative FL system.} \label{fig:model}
	\end{figure}
	In this paper, we investigate the design of the relay-assisted cooperative FL system. As shown in Fig. \ref{fig:model}, the system comprises an AP, a set of half-duplex relays, and a set of mobile devices. The devices upload the local model updates to the AP with the assistance of the relays. In particular, the relays adopt the amplify-and-forward (AF) strategy to assist the devices for over-the-air model aggregation in a cooperative manner. To the best of our knowledge, this is the first work that exploits the cooperative diversity of devices and relays for FL model aggregation. The main contributions of this paper are summarized as follows.
	\begin{itemize}
		\item We propose a novel two-phase AF-relay-assisted cooperative model aggregation framework to exploit the cooperative diversity and enhance the communication efficiency in FL systems. Specifically, in the first phase of local model uploading, the devices concurrently transmit their local model updates to both the relays and the AP. In the second phase, the devices and the relays simultaneously send the local model updates and the amplified received signals, respectively, to the AP. We then analytically derive the model aggregation error as a function of the transceiver design in the relay-assisted FL system.
		
		\item To improve the FL convergence, we investigate the joint optimization of the pre-equalization scalars at the devices/relays and the receive scalars at the AP to minimize the model aggregation error. The joint optimization problem is challenging because of the strong coupling among the variables. We propose an effective algorithm based on alternating optimization to alternately optimize the transmit scalars at the devices, the relaying scalars, and the receive scalars at the AP. This allows us to solve the problem in polynomial time by deriving analytical solutions of the relaying scalars and the receive scalars at the AP.
		
		\item We rigorously analyze the model aggregation MSE in a single-relay case. We show that inserting relays does not always lead to a better model aggregation accuracy. Furthermore, we derive conditions under which the relay-assisted FL scheme outperforms the conventional FL scheme without cooperative relaying. The analysis provides a useful guideline on how to deploy the relays in FL systems so that the relay assistance becomes beneficial.
	\end{itemize}
	Simulation results show that the proposed relay-assisted cooperative FL scheme achieves significant performance improvement compared with the existing model aggregation schemes. In particular, the proposed scheme achieves a smaller model aggregation error and higher learning accuracy, and performs very close to the ideal case with error-free communication channels. Moreover, the proposed scheme is more robust to the change of the transmit power, the locations, and the number of relays thanks to the cooperative diversity harvested from the relays and devices.
	
	The rest of this paper is organized as follows. In Section II, we describe the system model, the relay-assisted cooperative model aggregation framework, and the problem formulation. In Section III, we propose an alternating minimization method to solve the formulated problem. In Section IV, we analyze the performance of the proposed relay-assisted scheme. In Section V, we evaluate the proposed design via extensive simulations. Finally, we conclude the paper in Section VI.
	
	\emph{Notations}: We use regular letters, boldface lower-case letters, boldface upper-case letters, and calligraphy letters to denote scalars, vectors, matrices, and sets, respectively. The real and complex domains are denoted by $\mathbb{R}$ and $\mathbb{C}$, respectively. We use $\overline{(\cdot)}$, $(\cdot)^T$, and $(\cdot)^H$ to denote the conjugate, the transpose, and the conjugate transpose, respectively. We use $x_i$ to denote the $i$-th entry of vector $\mathbf{x}$, $x_{ij}$ to denote the $(i, j)$-th entry of matrix $\mathbf{X}$, $|\mathcal{S}|$ to denote the cardinality of set $\mathcal{S}$, $\text{diag}(\mathbf{x})$ to denote a diagonal matrix with the diagonal entries specified by $\mathbf{x}$, $\mathbf{I}_N$ to denote the $N \times N$ identity matrix, $\mathcal{CN}(\mu, \sigma^2)$ to denote the circularly-symmetric complex Gaussian distribution with mean $\mu$ and variance $\sigma^2$, and $\mathbb{E}[\cdot]$ to denote the expectation of a random variable.

	\section{System Model and Problem Formulation}
	\subsection{{System Model}}
	As shown in Fig. \ref{fig:model}, we consider a relay-assisted FL system, where $N$ relays are deployed to assist the communication between an AP and $K$ devices for FL. We assume that the AP, the devices, and the relays are all equipped with one antenna. Let $\mathcal{K} = \{1, \cdots, K\}$ denote the set of devices. The devices collaboratively train a shared AI model by using their local data. Let $\mathcal{D}_k  = \{(\mathbf{u}_{ki}, v_{ki}): 1 \leq i \leq D_k\}$ denote the local training dataset at device $k$ with $|\mathcal{D}_k| = D_k$, where $\mathbf{u}_{ki}$ is the input of training sample $i$ at device $k$ and $v_{ki}$ is the corresponding output. The learning objective is to minimize the following empirical loss function:
	\begin{equation}
		\min_{\mathbf{w} \in \mathbb{R}^{d \times 1}} F(\mathbf{w}) = \frac{1}{D} \sum_{k = 1}^K \sum_{i = 1}^{D_k} f(\mathbf{w}; \mathbf{u}_{ki}, v_{ki}),  \label{learning_objective}
	\end{equation}
	where $\mathbf{w}$ is the $d$-dimensional model parameter vector, $D = \sum_{k = 1}^K D_k$ is the total number of training samples of all devices, and $f(\mathbf{w}; \mathbf{u}_{ki}, v_{ki})$ is the loss function with respect to the training sample $(\mathbf{u}_{ki}, v_{ki})$. We can rewrite \eqref{learning_objective} as
	\begin{equation}
		\min_{\mathbf{w} \in \mathbb{R}^{d \times 1}} F(\mathbf{w}) = \sum_{k = 1}^K \rho_k F_k(\mathbf{w}; \mathcal{D}_k), \label{learning_objective2}
	\end{equation}
	where $\rho_k \triangleq \frac{D_k}{D}$ denotes the weight of device $k$, and
	\begin{equation}
		F_k(\mathbf{w}; \mathcal{D}_k) = \frac{1}{D_k} \sum_{(\mathbf{u}_{ki}, v_{ki}) \in \mathcal{D}_k} f(\mathbf{w}; \mathbf{u}_{ki}, v_{ki})
	\end{equation}
	denotes the local empirical loss function at device $k$.
	
	To solve the minimization problem in \eqref{learning_objective2}, we apply the canonical FL algorithm in \cite{mcmahan2017communication}, namely Federated Averaging (FedAvg), to optimize the model $\mathbf{w}$ iteratively through distributed on-device edge learning, which is coordinated by the AP through communication over wireless channels. Specifically, each learning iteration $t$, $1 \leq t \leq T$, consists of the following steps:
	\begin{enumerate}
		\item \emph{Model dissemination}: The AP broadcasts the updated global model $\mathbf{w}_t$ to all devices.
		
		\item \emph{Local model update}: Each device $k$ initializes $\mathbf{w}_{k, t}^{1} = \mathbf{w}_{t}$ and updates its local model by $\tau$-step batch gradient descent \cite{konevcny2016federated} using its local dataset $\mathcal{D}_k$. The updated local model is
		\begin{equation}
			\mathbf{w}_{k, t}^{\tau + 1} = \mathbf{w}_{t} - \lambda_t \sum_{i = 1}^{\tau} \nabla F_k(\mathbf{w}_{k, t}^{i}; \mathcal{D}_k), \label{local_model}
		\end{equation}
		where $\lambda_t$ is the learning rate, and $\nabla F_k(\mathbf{w}_{k, t}^{i}; \mathcal{D}_k) \in \mathbb{R}^{d \times 1}$ is the gradient of $F_k(\cdot)$ with respect to $\mathcal{D}_k$ at $\mathbf{w}_{k, t}^{i}$.
		
		\item \emph{Model aggregation}: Each device $k$ sends its local model change $\bm{\Delta}_{k, t} \triangleq \mathbf{w}_{k, t}^{\tau + 1} - \mathbf{w}_{t}$ to the AP through the wireless channels.\footnote{Alternatively, the devices can transmit either the local gradients or the local updated models to the AP. The proposed design and analysis in this paper are readily applicable to these two cases.} The AP estimates $\{\bm{\Delta}_{k, t}\}$ and aggregates these model changes to update the global model $\mathbf{w}_{t + 1}$ as
		\begin{equation}
			\mathbf{w}_{t + 1} = \mathbf{w}_{t} + \widehat{\sum_{k = 1}^K \rho_k \bm{\Delta}_{k, t}}, \label{global_model}
		\end{equation}
		where $\widehat{\sum_{k = 1}^K \rho_k \bm{\Delta}_{k, t}}$ is the estimate of the true weighted sum $\sum_{k = 1}^K \rho_k \bm{\Delta}_{k, t}$.
	\end{enumerate}
	
	Note that wireless fading and communication noise inevitably bring errors in uploading $\{\bm{\Delta}_{k, t}\}$ from the devices to the AP. Hence, the AP can only \emph{estimate} a global model change with additional errors, i.e., $\widehat{\sum_{k = 1}^K \rho_k \bm{\Delta}_{k, t}}$. In \eqref{global_model}, we use such an estimate instead of the true one to update $\mathbf{w}_{t + 1}$. The estimation accuracy of $\widehat{\sum_{k = 1}^K \rho_k \bm{\Delta}_{k, t}}$ critically affects the convergence of FL. In this paper, we shall propose a relay-assisted model aggregation scheme to improve the estimation accuracy by exploiting the cooperative diversity in relay networks. To facilitate our design on the relay operation, we first review the state-of-the-art model aggregation design without relays in the subsequent subsection.
	
	\subsection{Preliminaries on Over-the-Air Model Aggregation Without Relays}
	Over-the-air computation is an efficient scheme for model aggregation because it allows all devices to simultaneously upload their local model changes to the AP over the same radio resources. By exploiting the signal-superposition property of a wireless multiple-access channel, over-the-air model aggregation efficiently constructs the noisy estimate $\widehat{\sum_{k = 1}^K \rho_k \bm{\Delta}_{k, t}}$ at the AP. However, as detailed in the following, the performance of this scheme is limited by the devices with weak channels since all devices need to align the model signals at the AP with each other through channel pre-equalization.\footnote{We note that over-the-air model aggregation can also be achieved without transmitter-side signal alignment \cite{yang2021revisiting}. However, to mitigate the aggregation error and enhance the convergence rate, we consider the signal-alignment-based over-the-air model aggregation and employ relays to enhance its performance in this work.} Such limitation affects the convergence rate in FL and calls for effective solutions. This motivates the deployment of relays and our relay-assisted cooperative model aggregation design in the next subsection.
	
	To simplify the notation, we omit the iteration index $t$ in the sequel. Let $h_k \in \mathbb{C}$ denote the channel coefficient between device $k$ and the AP. We assume perfect channel state information (CSI) at the AP and the devices. Note that the entries in each local model change vector $\bm{\Delta}_k$ may vary significantly in values, which makes it difficult to effectively control the pre-equalization. To facilitate the pre-equalization design and transmit power control, each device $k$ first transforms the local model change vector $\bm{\Delta}_k \in \mathbb{R}^{d \times 1}$ into a normalized symbol vector $\mathbf{s}_k \in \mathbb{C}^{d \times 1}$ with zero mean and unit variance such that $\mathbb{E}[\mathbf{s}_k \mathbf{s}_{k'}^H] = \mathbf{0}$ and $\mathbb{E}[\mathbf{s}_k \mathbf{s}_k^H] = \mathbf{I}_d, \forall k, k' \in \mathcal{K}, k \neq k'$, where $\mathbf{0}$ is the all-zero matrix \cite{zhu2019broadband}. Specifically, each device $k$ computes the mean and variance of the local model parameter changes by
	\begin{equation}
		\Delta_{k, \text{mean}} = \frac{1}{d} \sum_{i = 1}^{d} \Delta_k[i],  ~~ \nu_k^2 = \frac{1}{d} \sum_{i = 1}^{d} (\Delta_k[i] - \Delta_{k, \text{mean}} )^2,
	\end{equation}
	where $\Delta_k[i]$ is the $i$-th entry of $\bm{\Delta}_k$. Then, the devices transmit the local statistics $\{\Delta_{k, \text{mean}}, \nu_k^2\}$ to the AP. Upon receiving $\{\Delta_{k, \text{mean}}, \nu_k^2\}$, the AP computes the global mean and variance by
	\begin{equation}
		\Delta_{\text{mean}} = \sum_{k = 1}^{K} \rho_k \Delta_{k, \text{mean}},  ~~~ \nu^2 = \sum_{k = 1}^{K} \rho_k \nu_k^2,
	\end{equation}
	and broadcasts $\Delta_{\text{mean}}$ and $\nu^2$ to the devices for normalization. For the $i$-th entry of $\bm{\Delta}_k$ at device $k$, the corresponding normalized symbol is given by
	\begin{equation}
		s_k[i] \triangleq \frac{\Delta_k[i] - \Delta_{\text{mean}}}{\nu}, ~~1 \leq i \leq d.
	\end{equation}
	
	To perform over-the-air model aggregation, $d$ communication rounds are required to upload the $d$-dimensional symbol vectors $\{\mathbf{s}_k, \forall k\}$ to the AP as the vectors are transmitted entry by entry. In each communication round $i$, $1 \leq i \leq d$, the devices concurrently upload $\{s_k[i], \forall k \}$ to the AP over the wireless channels after proper pre-equalization. The AP intends to estimate $x[i] \triangleq \sum_{k = 1}^K \rho_k s_k[i]$ from the received signal. We denote the estimate of $x[i]$ as $\hat{x}[i]$. After obtaining $\hat{x}[i]$, the AP can use the normalization factors $\Delta_{\text{mean}}$ and $\nu^2$ to retrieve the estimate of $\sum_{k = 1}^{K} \rho_k \Delta_k[i]$, i.e., $\widehat{\sum_{k = 1}^{K} \rho_k \Delta_k[i]}$, by a de-normalization step as
	\begin{equation}  \label{de-normalization}
		\widehat{\sum_{k = 1}^{K} \rho_k \Delta_k[i]} = \nu \hat{x}[i] + \Delta_{\text{mean}}.
	\end{equation}
	Following the above procedure, the AP obtains the desired estimate $\widehat{\sum_{k = 1}^K \rho_k \bm{\Delta}_{k}}$ after $d$ communication rounds to compute \eqref{global_model}. Due to the one-to-one mapping between the estimates $\widehat{\sum_{k = 1}^K \rho_k \Delta_k[i]}$ and $\hat{x}[i]$ in \eqref{de-normalization}, the estimation accuracy of $\widehat{\sum_{k = 1}^K \rho_k \Delta_k[i]}$ is directly determined by that of $\hat{x}[i]$.
	
	Let $a_k \in \mathbb{C}$ denote the complex-valued transmit scalar at device $k$. The received signal at the AP in the $i$-th communication round is given by
	\begin{equation}
		y[i] = \sum_{k = 1}^K h_k a_k s_k[i] + z[i],  \label{conventional_rec}
	\end{equation}
	where $z[i] \sim \mathcal{CN}(0, \sigma^2)$ is the additive white Gaussian noise (AWGN) at the AP. The AP estimates the weighted sum $x[i]$ from $y[i]$ by a de-noising receive scalar $c \in \mathbb{C}$ as
	\begin{equation}
		\hat{x}[i] = c y[i]= c \sum_{k = 1}^K h_k a_k s_k[i] + c z[i].  \label{conventional}
	\end{equation}
	We quantify the estimation accuracy by the MSE between $\hat{x}[i]$ and $x[i]$:
	\begin{equation}
		\bar{e}_{\text{no-relay}} \triangleq \mathbb{E} \left[|\hat x[i]-x[i]|^2 \right] = \sum_{k = 1}^K \left|c h_{k} a_{k} - \rho_k \right|^2  + |c|^2 \sigma^2. \label{conventional_mse}
	\end{equation}
	
	Here, we consider an individual transmit power constraint at each device
	\begin{equation}
		\mathbb{E} \left[|a_k s_k[i]|^2 \right]= |a_k|^2 \leq 2 P_0, ~~\forall k \in \mathcal{K},  \label{conventional_cons}
	\end{equation}
	where $2P_0$ is the maximum transmit power at each device.
	
	The minimum MSE of the model aggregation scheme in \eqref{conventional} is given in the following lemma.
	\begin{lemma} \label{lemmaA}
		With the power constraints in \eqref{conventional_cons}, the minimum $\bar{e}_{\text{no-relay}}$ is given by
		\begin{equation} \label{MMSE_A}
			\bar{e}_{\text{no-relay}} = \frac{\sigma^2}{2P_0} \max_{k \in \mathcal{K}} \frac{\rho_k^2}{|h_k|^2}.
		\end{equation}
	\end{lemma}
	\begin{IEEEproof}
		See Appendix \ref{appendixA}.
	\end{IEEEproof}
	
	Lemma \ref{lemmaA} implies that the model aggregation error is dominated by the device with the largest value of $\rho_k^2 / |h_k|^2$. The existing solutions \cite{yang2020federated, liu2020reconfigurable} alleviate severe aggregation distortion by excluding the devices with weak channel conditions. However, dropping those participants significantly decreases the number of exploited training data and hence slows down the learning convergence. Different from \cite{yang2020federated, liu2020reconfigurable}, we adopt cooperative relays to enhance the communication qualities of the devices with weak channels so that those devices do not have to be discarded. We describe the details in the next subsection.

	\subsection{Relay-Assisted Cooperative Model Aggregation}
	
	\begin{figure*}[t]
		\centering
		\includegraphics[scale=0.44]{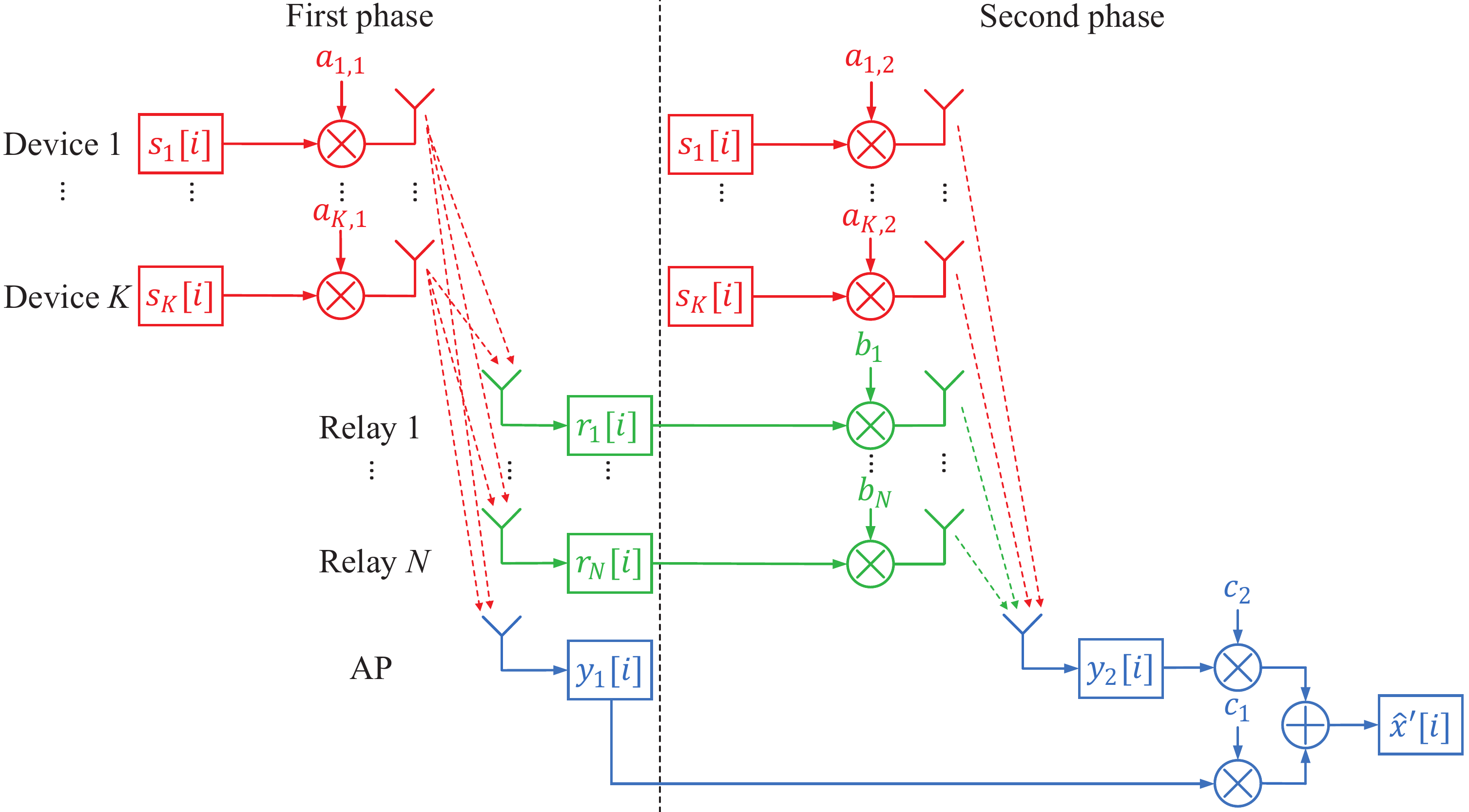}
		\caption{The two-phase relay-assisted scheme for aggregating $\hat{x}^\prime[i]$. In the first phase, the devices transmit the model signals to the relays and the AP. In the second phase, both the devices and the relays transmit signals to the AP.} \label{fig:diagram}
	\end{figure*}
	Let $\mathcal{N} = \{1, \cdots, N\}$ denote the set of relays, $g_{k, n} \in \mathbb{C}$ denote the channel coefficient between device $k$ and relay $n$, and $f_{n} \in \mathbb{C}$ denote the channel coefficient between relay $n$ and the AP. We assume perfect CSI at the relays. In the following, we propose a two-phase AF relaying design to enhance the over-the-air aggregation of $\{s_k[i]\}$ and construct the estimate of $\sum_{k = 1}^{K} \rho_k s_k[i]$ at the AP.
	
	Suppose that the relays are half-duplex, implying that they cannot transmit and receive at the same time. Correspondingly, we divide the local model uploading in each communication round into two phases. In the first phase, the devices concurrently transmit the model signals to both the relays and the AP over one radio resource block. In the second phase, the devices and the relays respectively send the model signals and the amplified received signals to the AP over another radio resource block. Since the transmission time of each symbol is fixed, the duration of each communication round in the proposed two-phase scheme is twice that in the scheme without relays in Section II-B. However, we will show in Section V that the proposed scheme still converges in a shorter time due to the relay assistance design. The details on the two phases of the $i$-th communication round are illustrated in Fig. \ref{fig:diagram}.
	
	In the first phase, all devices in $\mathcal{K}$ transmit their signals $\{s_k[i]\}$ to both the relays and the AP simultaneously after applying transmit scaling. Let $a_{k, 1} \in \mathbb{C}$ denote the complex-valued transmit scalar at device $k$ in the first phase. The received signal at relay $n$ is given by
	\begin{equation}  \label{rec_relay}
		r_n[i] = \sum_{k = 1}^K g_{k, n} a_{k, 1} s_k[i] + z_{r, n}[i],
	\end{equation}
	where $z_{r, n}[i] \sim \mathcal{CN}(0, \sigma^2)$ is the AWGN at relay $n$. Meanwhile, the received signal at the AP is given by
	\begin{equation}
		y_1[i] = \sum_{k = 1}^K h_{k} a_{k, 1} s_k[i] + z_1[i],
	\end{equation}
	where $z_1[i] \sim \mathcal{CN}(0, \sigma^2)$ is the AWGN at the AP in the first phase.
	
	In the second phase, each relay $n$ amplifies the received signal $r_n[i]$ by multiplying a complex-valued scalar $b_n \in \mathbb{C}$ and forwards it to the AP. The transmitted signal from relay $n$ is thus $b_n r_n[i]$. At the same time, each device $k$ retransmits $s_k[i]$ to the AP. We apply a transmit scalar $a_{k, 2} \in \mathbb{C}$ to $s_k[i]$ before the transmission. Then, the received signal at the AP is given by
	\begin{align}
		y_2[i] =& \sum_{n = 1}^N f_n b_n r_n[i] + \sum_{k = 1}^K h_{k} a_{k, 2} s_k[i] + z_2[i]  \nonumber \\
		=& \sum_{n = 1}^N f_n b_n \left(\sum_{k = 1}^K g_{k, n} a_{k, 1} s_k[i] + z_{r, n}[i]\right)  \nonumber\\
        &+ \sum_{k = 1}^K h_{k} a_{k, 2} s_k[i] + z_2[i],
	\end{align}
	where $z_2[i] \sim \mathcal{CN}(0, \sigma^2)$ is the AWGN at the AP in the second phase.
	
	Since each device transmits its signals in both phases, we equally allocate the maximum transmit power $2P_0$ to the two phases. The transmit power at each device is constrained by
	\begin{align}
		\mathbb{E} \left[|a_{k, 1} s_k[i]|^2 \right] = |a_{k, 1}|^2 \leq P_0, \ \forall k \in \mathcal{K},  \label{device_power} \\
		\mathbb{E} \left[|a_{k, 2} s_k[i]|^2 \right] = |a_{k, 2}|^2 \leq P_0, \ \forall k \in \mathcal{K}.
	\end{align}
	Let $P_r$ denote the maximum transmit power at each relay. Accordingly, the transmit power constraint at each relay is given by
	\begin{align}
		\mathbb{E} \left[|b_n r_n[i]|^2 \right] &= |b_n|^2 \left(\sum_{k = 1}^K |g_{k, n}|^2 |a_{k, 1}|^2 + \sigma^2 \right) \nonumber\\
        &\leq P_{r}, \;\forall n \in \mathcal{N}.
	\end{align}
	
	After receiving the signals $y_1[i]$ and $y_2[i]$ in the two phases, the AP applies two different de-noising receive scalars $c_1 \in \mathbb{C}$ and $c_2 \in \mathbb{C}$ to $y_1[i]$ and $y_2[i]$, respectively, and sums them up to obtain the desired estimate of the weighted sum $x[i] = \sum_{k = 1}^K \rho_k s_k[i]$ as $\hat{x}^\prime[i]$. That is,
	\begin{align} \label{x_estimate}
		&\hat{x}^\prime[i] = c_1 y_1[i] + c_2 y_2[i] \nonumber \\
		&~~~~= \sum_{k = 1}^K \Big(c_1 h_{k} a_{k, 1}  + c_2 h_{k} a_{k, 2} + c_2 a_{k, 1} \sum_{n = 1}^N f_n b_n g_{k, n} \Big) s_k[i]  \nonumber \\
		&~~~~~~~+ c_1 z_1[i] + c_2 z_2[i] + c_2 \sum_{n = 1}^N f_n b_n z_{r, n}[i].
	\end{align}
	Note that the relays are effective only when $c_2 \neq 0$. Therefore, we suppose $c_2 \neq 0$ hereafter. After estimating $\hat{\mathbf{x}}^\prime = [\hat{x}^\prime[1], \cdots, \hat{x}^\prime[d]]^T$, the AP can obtain a noisy estimate of $\sum_{k = 1}^K \rho_k \bm{\Delta}_{k}$ via the aforementioned de-normalization step, i.e., \eqref{de-normalization}, and update the global model by \eqref{global_model}. We refer to the above design as \emph{relay-assisted cooperative model aggregation} where the relays cooperatively help the devices align the signals at the AP. We calculate the MSE between $\hat{x}^\prime[i]$ and $x[i]$ as
    \begin{align} \label{MSE}
		\text{MSE}& \left(\hat{x}^\prime[i], x[i] \right) = \mathbb{E} \left[ \left|\hat{x}^\prime[i] - x[i] \right|^2 \right]  \nonumber \\
		=&\sum_{k = 1}^K \left|c_1 h_{k} a_{k, 1}  + c_2 h_{k} a_{k, 2} + c_2 a_{k, 1} \sum_{n = 1}^N f_n b_n g_{k, n} - \rho_k \right|^2  \nonumber \\
        &+ \left(|c_1|^2 + |c_2|^2  + |c_2|^2 \sum_{n = 1}^N |f_n|^2 |b_n|^2 \right) \sigma^2.
	\end{align}

	\subsection{Problem Formulation}
	As shown in \cite{liu2020reconfigurable}, the model aggregation MSE critically affects the FL performance in terms of both the convergence rate and optimality gap. In particular, a smaller MSE leads to a larger convergence rate and a smaller optimality gap. Motivated by this, we propose to minimize the MSE in \eqref{MSE} over the transmit scalars at the devices and relays $\{a_{k, 1}, a_{k, 2}, \forall k, b_n, \forall n\}$ and the de-noising receive scalars at the AP $\{c_1, c_2\}$. Specifically, we solve
	\begin{subequations} \label{problem1}
		\begin{align}
			&\min_{\{a_{k, 1}, a_{k, 2}, \forall k, b_n, \forall n, c_1, c_2\}}\; \sum_{k = 1}^K \Bigg|c_1 h_{k} a_{k, 1} + c_2 h_{k} a_{k, 2}  \nonumber \\
            &\quad\qquad\qquad\qquad + c_2 a_{k, 1} \sum_{n = 1}^N f_n b_n g_{k, n} - \rho_k \Bigg|^2  \nonumber \\
            &\quad\qquad\quad +\left(|c_1|^2 + |c_2|^2  + |c_2|^2 \sum_{n = 1}^N |f_n|^2 |b_n|^2 \right) \sigma^2  \label{obj1}  \\
			&\qquad\quad\  {\rm s.t.}\;
			|a_{k, 1}|^2 \leq P_0, \;\forall k \in \mathcal{K}, \label{cons1_1}  \\
			&\qquad\qquad\quad |a_{k, 2}|^2 \leq P_0, \;\forall k \in \mathcal{K}, \label{cons1_2}  \\
			&\qquad\qquad\quad |b_n|^2 \left(\sum_{k = 1}^K |g_{k, n}|^2 |a_{k, 1}|^2 + \sigma^2 \right) \leq P_{r}, \;\forall n \in \mathcal{N}.  \label{cons1_3}
		\end{align}
	\end{subequations}
	\begin{remark}
	 	The objective \eqref{obj1} is a linear combination of $K + 1$ terms, where the first $K$ terms quantify the misalignment errors of $K$ local models, and the last term represents the effective error induced by the communication noise in the two-phase model uploading scheme. Specifically, the $k$-th signal misalignment error depends on the transmit scalars of device $k$ (i.e., $a_{k, 1}$ and $a_{k, 2}$), the relaying scalars (i.e., $\{b_n\}$), and the two de-noising receive scalars at the AP (i.e., $c_1$ and $c_2$).
	\end{remark}

	Problem \eqref{problem1} is non-convex because of the coupling variables in both the objective function and the constraints in \eqref{cons1_3}. In the following section, we propose a low-complexity algorithm to effectively solve this problem.

	\section{Alternating Optimization for MSE Minimization}
	In this section, we propose an alternating minimization method to alternately optimize the transmit scalars at the devices, the transmit scalars at the relays, and the de-noising receive scalars at the AP.

	\subsection{Optimizing $\{a_{k, 1}, a_{k, 2}\}$ for Given $\{b_n, c_1, c_2\}$}	
	In this subsection, we optimize the transmit scalars at the devices, i.e., $\{a_{k, 1}, a_{k, 2}\}$, for given $\{b_n, c_1, c_2\}$. In this case, the MSE minimization problem in \eqref{problem1} is simplified as:
	\begin{align}  \label{problem_a}
		\min_{\{a_{k, 1}, a_{k, 2}\}}~&\sum_{k = 1}^K \Bigg|\left(c_1 h_{k} + c_2 \sum_{n = 1}^N f_n b_n g_{k, n} \right) a_{k, 1}  \nonumber\\
        &~~~~~~~+ c_2 h_{k} a_{k, 2}  - \rho_k \Bigg|^2  \\
		{\rm s.t.}~~~~~&\eqref{cons1_1}-\eqref{cons1_3} \nonumber.
	\end{align}
	Let $\theta_k = c_1 h_{k} + c_2 \sum_{n = 1}^N f_n b_n g_{k, n}$, and $\phi_k = c_2 h_{k}$. Define $\bm{\theta} \triangleq [\theta_1, \cdots, \theta_K]^T$, $\bm{\phi} \triangleq [\phi_1, \cdots, \phi_K]^T$, $\bm{a}_1 \triangleq [a_{1, 1}, \cdots, a_{K, 1}]^T$, $\bm{a}_2 \triangleq [a_{1, 2}, \cdots, a_{K, 2}]^T$, and $\bm{\rho} \triangleq [\rho_1, \cdots,$ $ \rho_K]^T$. Then, we can reformulate \eqref{problem_a} as
	\begin{align}  \label{problem1-3r}
		\min_{\bm{a}_1, \bm{a}_2}~&\left\| \text{diag}(\bm{\theta}) \bm{a}_1 + \text{diag}(\bm{\phi}) \bm{a}_2 - \bm{\rho} \right\|_2^2  \\
		{\rm s.t.}~~&\eqref{cons1_1}-\eqref{cons1_3} \nonumber.
	\end{align}
	Problem \eqref{problem1-3r} is jointly convex in $(\bm{a}_1, \bm{a}_2)$. Thus, we can apply a standard convex optimization solver, such as CVX \cite{cvx_solver}, to solve it.

	\subsection{Optimizing $\{b_n\}$ for Given $\{a_{k, 1}, a_{k, 2}, c_1, c_2\}$}
	For given $\{a_{k, 1}, a_{k, 2}, c_1, c_2\}$, we optimize the transmit scalars at the relays, i.e., $\{b_n\}$, by solving the following simplified MSE minimization problem:
	\begin{subequations} \label{problem_beta}	
		\begin{align}
			\min_{\{b_n\}}\;&\sum_{k = 1}^K \left|c_1 h_{k} a_{k, 1} + c_2 h_{k} a_{k, 2} + c_2 a_{k, 1} \sum_{n = 1}^N f_n b_n g_{k, n} - \rho_k \right|^2 \nonumber \\
            &+ |c_2|^2 \sum_{n = 1}^N |f_n|^2 |b_n|^2 \sigma^2  \label{obj_beta} \\
			{\rm s.t.}\;&|b_n|^2 \left(\sum_{k = 1}^K |g_{k, n}|^2 |a_{k, 1}|^2 + \sigma^2 \right) \leq P_{r},  \;\forall n \in \mathcal{N}. \label{cons_beta_1}
		\end{align}
	\end{subequations}
	Define $\mathbf{b} \triangleq [b_1, \cdots, b_N ]^T$, $\mathbf{f} \triangleq [f_1, \cdots, f_N  ]^T$, and $\mathbf{g}_k \triangleq [g_{k, 1}, \cdots, g_{k, N}  ]^T$. We recast Problem \eqref{problem_beta} as
	\begin{subequations} \label{problem_beta2}
		\begin{align}
			\min_{\mathbf{b}}\;&\sum_{k = 1}^K \left|c_2 a_{k, 1} \mathbf{g}_k^T \text{diag}\left(\mathbf{f} \right) \mathbf{b} + c_1 h_{k} a_{k, 1} + c_2 h_{k} a_{k, 2} - \rho_k \right|^2  \nonumber\\
            &+ |c_2|^2 \sigma^2 \mathbf{b}^H \text{diag}\left(\mathbf{f}^H \right) \text{diag}\left(\mathbf{f} \right) \mathbf{b}  \label{obj_beta2} \\
			{\rm s.t.}\;&|b_n|^2 \leq \hat{P}_{r, n},  ~~\forall n \in \mathcal{N}, \label{cons_beta2_1}
		\end{align}
	\end{subequations}
	where $\hat{P}_{r, n} \triangleq \frac{P_r}{\sum_{k = 1}^K |g_{k, n}|^2 |a_{k, 1}|^2 + \sigma^2}, \forall n \in \mathcal{N}$.
	
	Problem \eqref{problem_beta2} is a convex quadratically constrained quadratic programming (QCQP) problem, whose closed-form solution is given by the following proposition.
	\begin{proposition} \label{proposition_beta}
		Given $\{a_{k, 1}, a_{k, 2}, c_1, c_2\}$, the optimal $\mathbf{b}^*$ that minimizes the MSE in \eqref{problem_beta2} is given by
		\begin{equation}
			\mathbf{b}^* = P_{\mathcal{S}} \left(\hat{\mathbf{b}} \right),
		\end{equation}
		where
		\begin{align}
			\hat{\mathbf{b}} =& \left[c_2 \left(\sum_{k = 1}^K  |a_{k, 1}|^2 \overline{\mathbf{g}_k} \mathbf{g}_k^T + \sigma^2 \mathbf{I}_N \right) \text{diag}\left(\mathbf{f} \right) \right]^{-1}  \nonumber\\
            &\times \left(\sum_{k = 1}^K \left[\rho_k - h_k (c_1 a_{k, 1} + c_2 a_{k, 2})  \right] \overline{a_{k, 1} \mathbf{g}_k} \right),
		\end{align}
		$\mathcal{S} = \left\{|b_n|^2 \leq \hat{P}_{r, n}, \forall n \in \mathcal{N} \right\}$ is the convex feasible set of $\mathbf{b}$, and $P_{\mathcal{S}}: \mathbb{C}^{N \times 1} \rightarrow \mathbb{C}^{N \times 1}$ is the orthogonal projection operator and is applied element-wisely to $\hat{\mathbf{b}}$. Specifically, for relay $n$,
		\begin{align} \label{opt_beta}
			b_n^* = P_{\mathcal{S}} \left(\hat{b}_n \right) =
			\begin{cases}
				\frac{\hat{b}_n}{|\hat{b}_n|} \sqrt{\hat{P}_{r, n}} ,  &\mbox{if $|\hat{b}_n| > \sqrt{\hat{P}_{r, n}}$}, \\
				\hat{b}_n,  &\mbox{if $|\hat{b}_n| \leq \sqrt{\hat{P}_{r, n}}$}.
			\end{cases}
		\end{align}
	\end{proposition}
	\begin{IEEEproof}
		By setting the first-order derivative of \eqref{obj_beta2} with respect to $\mathbf{b}$ to $\mathbf{0}$, we have
		\begin{align}
			\mathbf{0} =& 2 \sum_{k = 1}^K \left(c_2 a_{k, 1} \mathbf{g}_k^T \text{diag}\left(\mathbf{f} \right) \mathbf{b} + c_1 h_{k} a_{k, 1} + c_2 h_{k} a_{k, 2} - \rho_k \right)  \nonumber \\
			&\times (c_2 a_{k, 1} \mathbf{g}_k^T \text{diag}\left(\mathbf{f} \right) )^{H} + 2 |c_2|^2 \sigma^2 \text{diag}\left(\mathbf{f}^H \right) \text{diag}\left(\mathbf{f} \right) \mathbf{b},
		\end{align}
		which yields
		\begin{align}
			\mathbf{b} =& \left[c_2 \left(\sum_{k = 1}^K  |a_{k, 1}|^2 \overline{\mathbf{g}_k} \mathbf{g}_k^T + \sigma^2 \mathbf{I}_N \right) \text{diag}\left(\mathbf{f} \right) \right]^{-1}  \nonumber\\
            &\times \left(\sum_{k = 1}^K \left[\rho_k - h_k (c_1 a_{k, 1} + c_2 a_{k, 2})  \right] \overline{a_{k, 1} \mathbf{g}_k} \right).
		\end{align}
		Considering the feasibility of each $b_n$, the projection operation is needed to project each $b_n$ on its feasible set. Due to the optimality of the orthogonal projection on a convex set \cite{cvx}, we can obtain the expression of the optimal $b_n^*$ as in \eqref{opt_beta}, which completes the proof.
	\end{IEEEproof}

	\subsection{Optimizing $c_1$ for Given $\{a_{k, 1}, a_{k, 2}, b_n, c_2\}$}
	For given $\{a_{k, 1}, a_{k, 2}, b_n, c_2\}$, we optimize the de-noising receive scalar $c_1$ at the AP by solving the following simplified MSE minimization problem:
	\begin{align}
		\min_{c_1}&\sum_{k = 1}^K \left|c_1 h_{k} a_{k, 1}  + c_2 \Bigg(h_{k} a_{k, 2} + a_{k, 1} \sum_{n = 1}^N f_n b_n g_{k, n} \Bigg) - \rho_k \right|^2  \nonumber\\
        &+ |c_1|^2 \sigma^2.  \label{problem_eta1}
	\end{align}
	Problem \eqref{problem_eta1} is an unconstrained quadratic programming (QP) problem. The optimal solution is given in the following proposition.
	\begin{proposition} \label{proposition_eta1}
		Given $\{a_{k, 1}, a_{k, 2}, b_n, c_2\}$, the optimal $c_1^*$ that minimizes the MSE in \eqref{problem_eta1} is given by
		\begin{equation} \label{opt_eta1}
			c_1^* = \frac{\sum\limits_{k = 1}^K \left[\rho_k - c_2 \left(h_{k} a_{k, 2} + a_{k, 1} \sum\limits_{n = 1}^N f_n b_n g_{k, n} \right) \right] \overline{h_{k} a_{k, 1}} }{\sum\limits_{k = 1}^K |h_{k}|^2 |a_{k, 1}|^2 + \sigma^2}.
		\end{equation}
	\end{proposition}
	\begin{IEEEproof}
		By setting the first-order derivative of \eqref{problem_eta1} with respect to $c_1$ as $0$, we have
		\begin{align} \label{deri_eta1}
			2 \sum_{k = 1}^K& \left[c_1 h_{k} a_{k, 1}  + c_2 \left(h_{k} a_{k, 2} + a_{k, 1} \sum_{n = 1}^N f_n b_n g_{k, n} \right) - \rho_k \right]  \nonumber\\
        &\times \overline{h_{k} a_{k, 1}} + 2 c_1 \sigma^2 = 0.
		\end{align}
		Note that \eqref{deri_eta1} can be rearranged as
		\begin{equation}
			c_1 = \frac{\sum\limits_{k = 1}^K \left[\rho_k - c_2 \left(h_{k} a_{k, 2} + a_{k, 1} \sum\limits_{n = 1}^N f_n b_n g_{k, n} \right) \right] \overline{h_{k} a_{k, 1}}}{\sum\limits_{k = 1}^K |h_{k}|^2 |a_{k, 1}|^2 + \sigma^2},
		\end{equation}
		which completes the proof.
	\end{IEEEproof}

	\subsection{Optimizing $c_2$ for Given $\{a_{k, 1}, a_{k, 2}, b_n, c_1\}$}
	In this subsection, we optimize the de-noising receive scalar $c_2$ at the AP for given $\{a_{k, 1}, a_{k, 2},$ $b_n, c_1\}$. Specifically, we solve the following MSE minimization problem:
	\begin{align}
		\min_{c_2}&\sum_{k = 1}^K \left|c_1 h_{k} a_{k, 1}  + c_2 \Bigg(h_{k} a_{k, 2} + a_{k, 1} \sum_{n = 1}^N f_n b_n g_{k, n} \Bigg) - \rho_k \right|^2  \nonumber\\
        &+ |c_2|^2 \left(1 + \sum_{n = 1}^N |f_n|^2 |b_n|^2 \right) \sigma^2.  \label{problem_eta2}
	\end{align}
	Similar to \eqref{problem_eta1}, the optimal solution to \eqref{problem_eta2} is given in the following proposition.
	\begin{proposition} \label{proposition_eta2}
		Given $\{a_{k, 1}, a_{k, 2}, b_n, c_1\}$, the optimal $c_2^*$ that minimizes the MSE in \eqref{problem_eta2} is given by \eqref{opt_eta2},  shown at the top of the next page.
		\begin{figure*}
            \begin{equation}  \label{opt_eta2}
    			c_2^* = \frac{\sum_{k = 1}^K (\rho_k - c_1 h_{k} a_{k, 1}) \overline{\left(h_{k} a_{k, 2} + a_{k, 1} \sum_{n = 1}^N f_n b_n g_{k, n} \right)} }{\sum_{k = 1}^K \left|h_{k} a_{k, 2} + a_{k, 1} \sum_{n = 1}^N f_n b_n g_{k, n} \right|^2 + \left(1 + \sum_{n = 1}^N |f_n|^2 |b_n|^2 \right) \sigma^2}.
    		\end{equation}
        \end{figure*}
	\end{proposition}
	\begin{IEEEproof}
		By setting the first-order derivative of \eqref{problem_eta2} with respect to $c_2$ as $0$, we have
		\begin{align} \label{deri_eta2}
			2 \sum_{k = 1}^K& \left[c_1 h_{k} a_{k, 1}  + c_2 \left(h_{k} a_{k, 2} + a_{k, 1} \sum_{n = 1}^N f_n b_n g_{k, n} \right) - \rho_k \right]  \nonumber  \\
			&\times \overline{\left(h_{k} a_{k, 2} + a_{k, 1} \sum_{n = 1}^N f_n b_n g_{k, n} \right)}  \nonumber\\
            &+ 2 c_2 \left(1 + \sum_{n = 1}^N |f_n|^2 |b_n|^2 \right) \sigma^2 = 0.
		\end{align}
		Note that \eqref{deri_eta2} can be rearranged as in \eqref{opt_eta2_proof}, shown at the top of the next page,
        \begin{figure*}
    		\begin{equation} \label{opt_eta2_proof}
    			c_2 = \frac{\sum_{k = 1}^K (\rho_k - c_1 h_{k} a_{k, 1}) \overline{\left(h_{k} a_{k, 2} + a_{k, 1} \sum_{n = 1}^N f_n b_n g_{k, n} \right)} }{\sum_{k = 1}^K \left| h_{k} a_{k, 2} + a_{k, 1} \sum_{n = 1}^N f_n b_n g_{k, n} \right|^2 + \left(1 + \sum_{n = 1}^N |f_n|^2 |b_n|^2 \right) \sigma^2}.
    		\end{equation}
            \hrulefill
        \end{figure*}
		which completes the proof.
	\end{IEEEproof}

	The overall alternating minimization algorithm for solving Problem \eqref{problem1} is summarized in Algorithm 1. We initialize $\Big\{a_{k, 1}^{(0)}, a_{k, 2}^{(0)}, \forall k, c_1^{(0)}, c_2^{(0)} \Big\}$ based on the channel inversion power control and initialize $\big\{b_n^{(0)}, \forall n \big\}$ such that the constraints in \eqref{cons1_3} are all active. The proposed algorithm terminates when either the maximum number of iterations $J_{\text{max}}$ is reached or the improvement on the objective value is less than a predetermined threshold $\epsilon$. Since the objective value of Problem \eqref{problem1} is non-increasing after each optimization step, Algorithm 1 is guaranteed to converge to a stationary point of Problem (23) \cite{nonlinear}.

	\subsection{Computational Complexity}
	When solving \eqref{problem1-3r}, CVX adopts the interior point method. As such, the complexity of Line 5 of Algorithm 1 is upper bounded by $O(K^3)$. The complexity of Line 6 is bounded by $O(N^3)$ because the calculation in \eqref{opt_beta} involves the inversion of an $N \times N$ matrix. In addition, the complexity of Lines 7--8 is $O(1)$, with the help of the closed-form expressions \eqref{opt_eta1} and \eqref{opt_eta2}. As a result, the overall complexity of the proposed alternating minimization method is upper bounded by $O \left(J_{\text{max}} \left(K^3 + N^3 \right) \right)$, implying that it can solve Problem \eqref{problem1} in polynomial time.

	\section{Performance Analysis on the Single-Relay Scenario}
	In this section, we analyze the model aggregation accuracy achieved by the proposed relay-assisted cooperative FL scheme. To simplify the derivation, we here consider that only one relay is deployed in the system. Since we assume $N = 1$, the relay index $n$ is omitted throughout this section. We use $f \in \mathbb{C}$ and $g_k \in \mathbb{C}$ to denote the channel coefficients from the relay to the AP and from device $k$ to the relay, respectively. Moreover, we assume an equal size of local training data for all devices. In other words, the aggregation weights $\rho_k = \rho, \forall k$, with $\rho \triangleq 1 / K$.
	
	Our analysis reports that inserting a relay, in general, does not necessarily give rise to performance improvement for over-the-air model aggregation. This is because the received signal at the relay in the first phase contains the communication noise in the device-relay channels; see \eqref{rec_relay}. This noise is coupled with the transmitted signals in the second phase as it is amplified together with the model signals, which potentially distorts the signal alignment at the AP. As detailed in the sequel, we show that the proposed single-relay-assisted cooperative FL achieves a smaller MSE than the one without relays, given that the transmit power at the relay is sufficiently large and the relay channels are sufficiently strong.
	
	To begin our analysis, recall from Lemma \ref{lemmaA} that with $\rho_k=\rho$, the model aggregation error of FL without relays is given by
	\begin{equation}  \label{xxxx}
		\bar{e}_{\text{no-relay}}  = \frac{\rho^2 \sigma^2}{2P_0} \max_{k \in \mathcal{K}} \frac{1}{|h_k|^2}.
	\end{equation}
	We shall compare the achievable MSE of our formulated problem in \eqref{problem1} with \eqref{xxxx}.
	For the single-relay system, the original problem in \eqref{problem1} is simplified as
	\begin{subequations} \label{problem_single}
		\begin{align}
			\min_{\{a_{k, 1}, a_{k, 2}, b, c_1, c_2\}}&\sum_{k = 1}^K \left|c_1 h_{k} a_{k, 1} + c_2 f b g_{k} a_{k, 1} + c_2 h_{k} a_{k, 2} - \rho \right|^2 \nonumber \\
            &+ \left(|c_1|^2 + |c_2|^2  + |c_2|^2 |f|^2 |b|^2 \right) \sigma^2  \label{obj_single}  \\
			{\rm s.t.}~~~~~~
			&\eqref{cons1_1}, \eqref{cons1_2},  \nonumber \\
			&|b|^2 \left(\sum_{k = 1}^K |g_{k}|^2 |a_{k, 1}|^2 + \sigma^2 \right) \leq P_{r},  \label{cons_single}
		\end{align}
	\end{subequations}
	where $b$ is the transmit scalar at the single relay and \eqref{cons_single} is the corresponding transmit power constraint. The strong coupling of the optimization variables $\{a_{k, 1}, a_{k, 2}, b, c_1, c_2\}$ in the objective function \eqref{obj_single} makes it difficult to directly analyze the optimal value of \eqref{problem_single}. Therefore, we analyze a \emph{sub-optimal} solution to \eqref{problem_single} by restricting $c_1 = 0$. In other words, we analyze the upper bound of the minimum value of \eqref{obj_single} and study when this bound is less than $\bar{e}_{\text{no-relay}}$, which is given in \eqref{xxxx}. Let $\text{SNR}_k \triangleq P_0 |h_k|^2 / \sigma^2$ be the maximum received signal-to-noise ratio (SNR) for the transmission from device $k$ to the AP in the first phase. Similarly, for our proposed cooperative scheme in Section II-C, let $\text{SNR}_{k, \text{relay}} \triangleq P_0 |g_k|^2 / \sigma^2$ denote the maximum received SNR for the transmission from device $k$ to the relay in the first phase, and $\text{SNR}_{\text{relay-AP}} \triangleq P_r |f|^2 / \sigma^2$ denote the maximum received SNR for the transmission from the relay to the AP in the second phase. Define $\bar{e}$ as the minimum MSE achieved by solving \eqref{problem_single}, and $\delta \triangleq \frac{\min_{k \in \mathcal{K}} \text{SNR}_k}{\min_{k \in \mathcal{K}} \text{SNR}_{k, \text{relay}}} \geq 0$. We derive the following theorem to compare the model aggregation errors with and without relays (i.e., $\bar{e}$ and $\bar{e}_{\text{no-relay}}$).
	\begin{theorem} \label{theorem1}
		Suppose $N=1$ and $\rho_k=\rho,\forall k$. We have
		\begin{equation}
			\bar{e} \leq \bar{e}_{\text{no-relay}}  \label{MSE_e}
		\end{equation}
		if the following two conditions hold:
		\begin{align}
			\delta	&\leq 1,  \label{snr_relay} \\
			\text{SNR}_{\text{relay-AP}}& \geq \frac{ K \min_{k \in \mathcal{K}} \text{SNR}_{k} + \delta }{ \big(1 + \sqrt{2 - 2 \delta}	\big)^2 }.	\label{snr_final}
		\end{align}
	\end{theorem}
	\begin{IEEEproof}
		See Appendix \ref{appendixC}.
	\end{IEEEproof}

	\begin{algorithm}[t]  \label{algorithm1}
		\caption{Alternating Minimization Method for Solving Problem \eqref{problem1}.}
		\begin{algorithmic}[1]
			\STATE \textbf{Input:} $J_{\text{max}}$ and $\epsilon$.
			\STATE \textbf{Initialization:} $a_{k, 1}^{(0)} = a_{k, 2}^{(0)} = \frac{\sqrt{P_0}  \rho_k}{h_k \max_{k \in \mathcal{K}} \frac{\rho_k}{|h_k|}}, \forall k \in \mathcal{K}$; $b_n^{(0)} = \sqrt{\frac{P_r}{\sum_{k = 1}^K |g_{k, n}|^2 |a_{k, 1}^{(0)} |^2 + \sigma^2}},  \forall n \in \mathcal{N}$; $c_1^{(0)} = c_2^{(0)} = \frac{1}{2 \sqrt{P_0}} \max_{k \in \mathcal{K}} \frac{\rho_k}{|h_k|}$.
			\STATE Compute $\text{MSE}^{(0)}$ by substituting $\big\{a_{k, 1}^{(0)}, a_{k, 2}^{(0)}, b_n^{(0)}, c_1^{(0)}, $ $c_2^{(0)} \big\}$ into \eqref{obj1};
			\FOR{$j = 1, 2, \cdots, J_{\text{max}}$}
			\STATE Compute $\{a_{k, 1}^{(j)}, a_{k, 2}^{(j)}\}$ by solving \eqref{problem1-3r};
			\STATE Compute $\{b_n^{(j)}\}$ by \eqref{opt_beta};
			\STATE Compute $c_1^{(j)}$ by \eqref{opt_eta1};
			\STATE Compute $c_2^{(j)}$ by \eqref{opt_eta2};
			\STATE Update $\text{MSE}^{(j)}$ by substituting $\big\{a_{k, 1}^{(j)}, a_{k, 2}^{(j)}, b_n^{(j)}, c_1^{(j)}, $ $c_2^{(j)} \big\}$ into \eqref{obj1};
			\IF{$\frac{|\text{MSE}^{(j)} - \text{MSE}^{(j - 1)}|}{|\text{MSE}^{(j)}|} \leq \epsilon$}
			\STATE Early stop;
			\ENDIF
			\ENDFOR
			\STATE \textbf{Output:} $\left\{a_{k, 1}^{(j)}, a_{k, 2}^{(j)}, b_n^{(j)}, c_1^{(j)}, c_2^{(j)} \right\}$.
		\end{algorithmic}
	\end{algorithm}

	Theorem \ref{theorem1} implies that the relay-assisted scheme achieves a smaller model aggregation error than the scheme without relays when the conditions in \eqref{snr_relay} and \eqref{snr_final} are satisfied. In particular, the condition in \eqref{snr_relay} requires the channel condition of the worst device-relay link to be no worse than that of the worst device-AP link. On the other hand, the condition in \eqref{snr_final} requires sufficiently large relay transmit power and/or strong enough relay-AP channel.
	
	Here, we explain the meaning of the sufficient conditions in \eqref{snr_relay} and \eqref{snr_final}. On one hand, to ensure that the relay assistance is beneficial, the model aggregation at the relay should be more accurate than that at the AP. Otherwise, since the additional noise raised at the relay (i.e., $z_{r,n}[i]$ in \eqref{rec_relay}) is also amplified and forwarded to the AP, this noise will damage the final aggregation accuracy at the AP compared with the relaying-free aggregation scheme in \eqref{conventional}. Notice that $\bar{e}_{\text{no-relay}}$ in \eqref{xxxx} is proportional to $\max_{k \in \mathcal{K}} 1/|h_k|^2$, or equivalently $1 /( \min_{k \in \mathcal{K}} |h_k|^2)$, showing that the direct over-the-air aggregation error is determined by the device with the minimum channel gain. Similarly, considering the transmission from the devices to the relay in \eqref{rec_relay}, the aggregation error is inversely proportional to the minimum channel gain $\min_{k \in \mathcal{K}} |g_k|^2$. In \eqref{snr_relay}, we guarantee that $\min_{k \in \mathcal{K}} |g_k|^2 \geq \min_{k \in \mathcal{K}} |h_k|^2$, or equivalently, $\min_{k \in \mathcal{K}} \text{SNR}_{k, \text{relay}} \geq \min_{k \in \mathcal{K}} \text{SNR}_k$. That is, when the transmitted signals are received by either the relay or the AP, the minimum received SNR at the relay from the devices should be no less than that at the AP. On the other hand, the condition in \eqref{snr_final} guarantees that in the second phase the signal transmitted from the relay to the AP achieves a sufficiently high SNR so that the additional noise in the AF operation does not impair the final model aggregation. The combination of these two conditions assures that the relay does improve the model aggregation accuracy in our design. Note that the relay transmit power is typically larger than the device transmit power. Therefore, the conditions in \eqref{snr_relay} and \eqref{snr_final} can be easily met by placing the relay at an intermediate position between the devices and the AP so that the average channel condition of each relaying link is better than that of the corresponding direct link.
	
	We highlight that the value of $\delta$ in Theorem 4.1 reflects the effect of the relay location and characterizes the tradeoff between the two hops of device-relay-AP links. Specifically, a larger $\delta$ is equivalent to a smaller $\min_{k \in \mathcal{K}} \text{SNR}_{k, \text{relay}}$ given fixed $\{\text{SNR}_k, \forall k\}$, which corresponds to worse device-relay channels. On the other hand, we observe from \eqref{snr_final} that the required lower bound on $\text{SNR}_{\text{relay-AP}}$ increases with $\delta$, and hence a larger $\delta$ corresponds to a stricter condition on the achievable SNR for the relay-AP channel. Therefore, Theorem \ref{theorem1} implies that the worse device-relay channels we have in the first hop, the higher SNR we need for the relay-AP channel in the second hop to guarantee $\bar{e} \leq \bar{e}_{\text{no-relay}}$. With respect to $\delta$, the relay should be placed properly to balance the SNRs of the two hops of the relaying links (i.e., $\min_{k \in \mathcal{K}} \text{SNR}_{k, \text{relay}}$ and $\text{SNR}_{\text{relay-AP}}$).

	When $\delta$ achieves its maximum value $1$, it means that the relay achieves as high model aggregation accuracy as the direct aggregation at the AP in the first phase. We here provide a more intuitive explanation on Theorem 4.1 in this special case. With $\delta=1$, we have $\min_{k \in \mathcal{K}} \text{SNR}_{k, \text{relay}} = \min_{k \in \mathcal{K}} \text{SNR}_k$. Furthermore, the condition in \eqref{snr_final} becomes
	\begin{align}  \label{snr_limit}
		\text{SNR}_{\text{relay-AP}} &\geq K \min\nolimits_{k \in \mathcal{K}} \text{SNR}_{k} + 1
        \nonumber\\
        &= K \min\nolimits_{k \in \mathcal{K}} \text{SNR}_{k, \text{relay}} + 1  \nonumber\\
        &= \frac{1}{\sigma^2} \Big( K P_0 \min\nolimits_{k \in \mathcal{K}} |g_k|^2 + \sigma^2 \Big).
	\end{align}
	The last term, $ ( K P_0 \min_{k \in \mathcal{K}} |g_k|^2 + \sigma^2 )/\sigma^2$, is an SNR term, where the signal is taken as the received signal at the relay $r_n[i]$ given in \eqref{rec_relay}. We refer to this SNR as the effective received SNR at the relay. The condition in \eqref{snr_limit} implies that when $\delta=1$, the relay-assisted scheme outperforms the scheme without relaying if the relay forwards its received model information to the AP with a higher SNR than its effective received SNR.

	\begin{remark}
		We emphasize that the conditions in \eqref{snr_relay} and \eqref{snr_final} are sufficient but not necessary for the single-relay case. We show in Section V through numerical results that the proposed scheme may still outperform the one without relays even if the conditions in Theorem \ref{theorem1} are violated. Moreover, our numerical results in Section V find that the model aggregation accuracy can be further enhanced by introducing more relays because increasing the number of relays further increases the cooperative diversity \cite{laneman2003distributed}.
	\end{remark}

	\begin{table}[t]
		\caption{System Parameters}
		\centering
		\begin{tabular}{|c|c|c|c|c|c|} 
			\hline
			Parameter &Value &Parameter &Value &Parameter &Value \\
			\hline
			$K$ &20 &$J_{\text{max}}$ &100 &$\epsilon$ &$10^{-4}$ \\
			\hline
			$P_0$ &0.05 W &$P_r$ &0.1 W &$\sigma^2$ &$-70$ dBm \\
			\hline
			$G_A$ &4.11 &$PL$ &3 &$f_c$ &915 MHz \\
			\hline
		\end{tabular}
		\label{table:parameter}
	\end{table}	
	\section{Numerical Results}
	In this section, we conduct numerical experiments to evaluate the performance of the proposed relay-assisted cooperative FL design. The simulation setup is presented in Section V-A. Simulation codes of our algorithm are available at https://github.com/zhlinup/Relay-FL.

    \begin{table}[t]
		\centering
		\caption{Average NMSE of each scheme with $\sigma^2 = -100 \textnormal{ dBm}$}
		\begin{tabular}{|c|c|}
			\hline
			Scheme & NMSE (dB) \\
			\hline
			Error-free channel & $-\infty$ \\
			\hline
			Proposed scheme & $-37.2380$ \\
			\hline
			FL without relays [32] & $-25.2075$ \\
			\hline
			Relay-assisted scheme in [23] & $-23.9929$ \\
			\hline
		\end{tabular}
		\label{table:NMSE_SNR130}
	\end{table}

    \begin{figure}[t]
		\centering
		\includegraphics[width=0.95\linewidth]{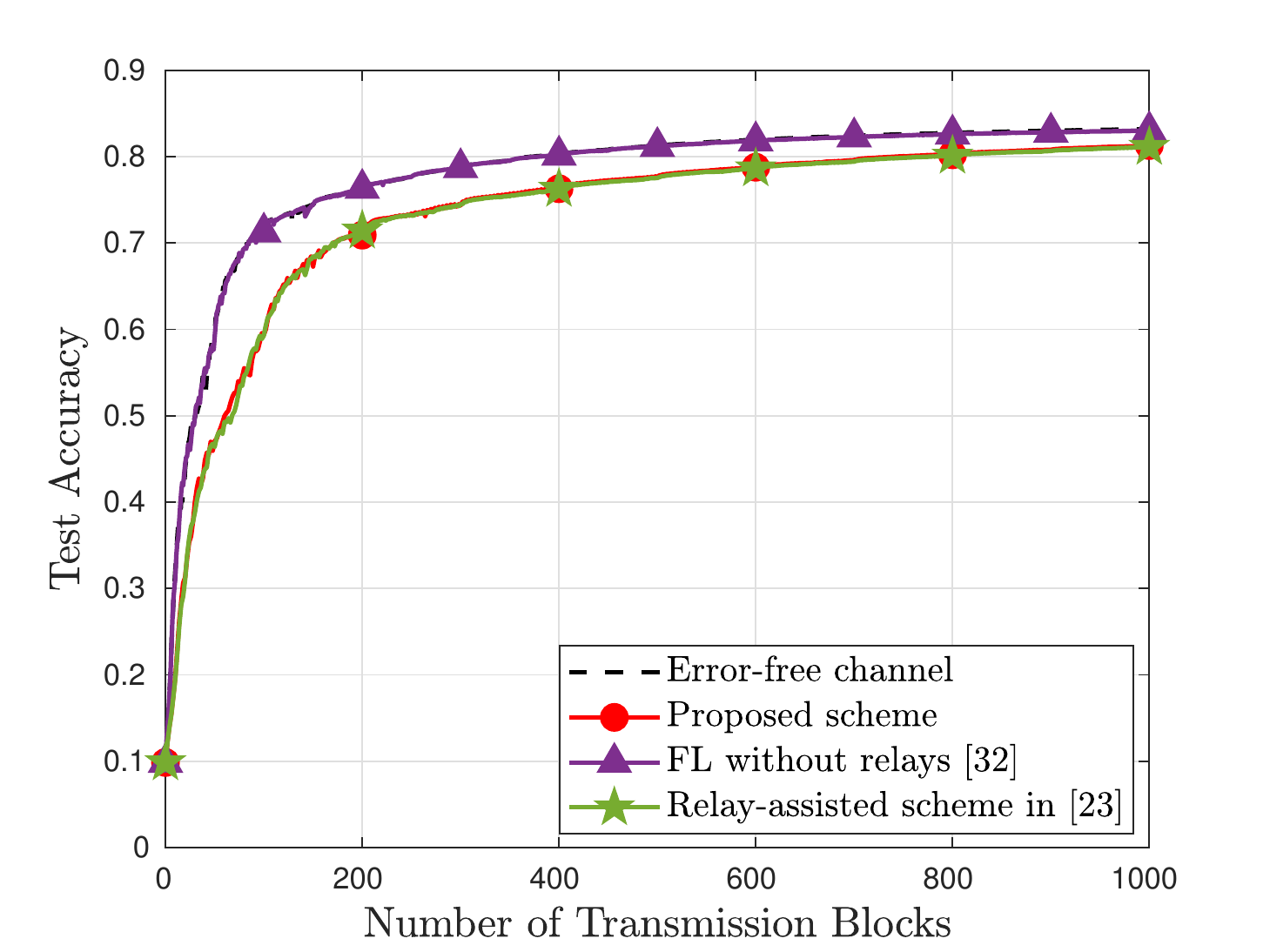}
		\caption{Test accuracy with $N = 1, P_r = 0.1$ W, $\sigma^2 = -100$ dBm.} \label{fig:cmp_time_N1_SNR130}
	\end{figure}
    \begin{table}[t]
		\centering
		\caption{Average NMSE of each scheme with $\sigma^2 = -70 \textnormal{ dBm}$}
		\begin{tabular}{|c|c|}
			\hline
			Scheme & NMSE (dB) \\
			\hline
			Error-free channel & $-\infty$ \\
			\hline
			Proposed scheme & $-6.2902$ \\
			\hline
			FL without relays [32] & $4.7907$ \\
			\hline
			Relay-assisted scheme in [23] & $-1.6358$ \\
			\hline
		\end{tabular}
		\label{table:NMSE}
	\end{table}

    \begin{figure}[t]
		\centering
		\includegraphics[width=0.95\linewidth]{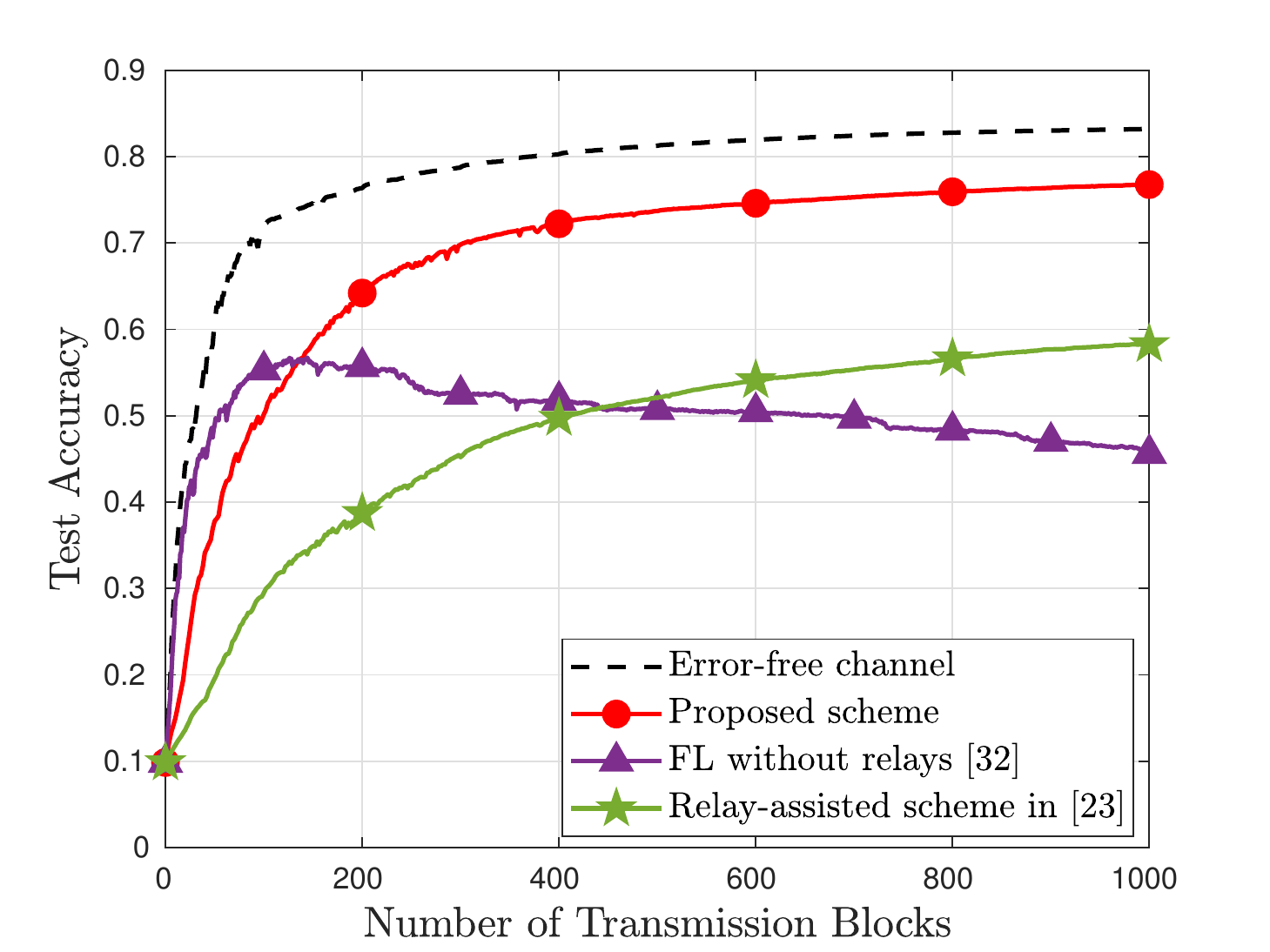}
		\caption{Test accuracy with $N = 1, P_r = 0.1$ W, $\sigma^2 = -70$ dBm.} \label{fig:cmp_time_N1}
	\end{figure}
	\subsection{Simulation Setup}
	We consider a Rayleigh fading channel model where the channel coefficients are given by the small-scale fading coefficients multiplied by the square root of the path loss. Specifically, the small-scale fading coefficients follow the standard independent and identically distributed (i.i.d.) Gaussian distribution. The path loss of the channel between any two nodes $i$ and $j$, $\forall i, j \in \{\text{AP}, \mathcal{N}, \mathcal{K}\}$, follows the free-space path loss model $G_A \Big(\frac{3 \cdot 10^8}{4 \pi f_c d_{i, j}} \Big)^{PL}$, where $G_A$ denotes the antenna gain, $f_c$ denotes the carrier frequency, $d_{i, j}$ denotes the distance between nodes $i$ and $j$, and $PL$ denotes the path loss exponent. The locations of the nodes will be specified later. Unless otherwise stated, the values of the system parameters are listed in Table \ref{table:parameter}.

	We simulate a learning task that classifies 10 classes of fashion products over the Fashion-MNIST dataset \cite{FashionMNIST}. The dataset consists of 60,000 training samples and 10,000 test samples, with each sample being a $28 \times 28$ grayscale image. The local training samples are drawn from the training set following an i.i.d. uniform distribution, with $D_k = \left\lfloor \frac{60000}{K} \right\rfloor$ for each device $k$. The learning model is a convolutional neural network that consists of two $5 \times 5$ convolutional layers (each followed by $2 \times 2$ max pooling), a batch normalization layer, a fully connected layer with $50$ neurons and Relu activation, and a softmax output layer (total number of parameters $d = 21921$). The loss function $F(\cdot)$ is the cross-entropy loss. The learning rate $\lambda_t$ is set as $\lambda_t = \max \big\{0.05 \cdot 0.9^{\left\lfloor \frac{t}{50} \right\rfloor} ,10^{-5} \big\}$. For ease of exposition, we assume that each device performs a single-step batch gradient descent in each learning iteration, i.e., $\tau = 1$, and focus on the communication design of relay-assisted cooperative model aggregation. The FL performance is measured by the test accuracy, which is defined as $\frac{\text{the number of correctly classified test images}}{10000} \in [0, 1]$. The numerical results in this section are averaged over 50 Monte Carlo trials. For comparison, we consider the following baselines:
	\begin{itemize}
		\item Error-free channel: The local model changes are assumed to be transmitted through noiseless channels, i.e., $\sigma^2 = 0$. The AP receives all local model changes without communication error and updates the global model by \eqref{global_model} with $\widehat{\sum_{k = 1}^K \rho_k \bm{\Delta}_{k}} = \sum_{k = 1}^K \rho_k \bm{\Delta}_{k}$. This ideal benchmark characterizes the best possible FL performance.
		
		\item Conventional scheme without relays \cite{chen2018uniform}: The devices transmit the local model changes to the AP by following \eqref{conventional_rec}--\eqref{conventional} under the constraints in \eqref{conventional_cons}. This baseline achieves the communication MSE given in \eqref{MMSE_A}.
		
		\item The method in \cite{wang2020optimized}: The devices transmit the local model changes to the AP through the $N$ relays following a two-phase AF relaying protocol. Different from the proposed scheme in this paper, the devices here do not retransmit in the second phase (i.e., $a_{k, 2} = 0, \forall k$), and the AP does not receive in the first phase (i.e., $c_1 = 0$). Similarly to \eqref{conventional_cons}, the individual maximum transmit power of each device is $2 P_0$.
	\end{itemize}

	For the error-free benchmark and the FL scheme without relays (i.e., the scheme in \cite{chen2018uniform}), we need $d$ time slots for transmitting the local model changes and updating \eqref{global_model} in each learning round. In contrast, the relay-assisted FL methods (i.e., our method and the one in \cite{wang2020optimized}) double the required communication time because of the half-duplex nature of the relays; see Section II-C. In the following simulations, we refer to $d$ time slots as one transmission block and control the total number of transmission blocks to be the same for all the algorithms. In other words, the total number of learning iterations $T$ of the error-free benchmark and the baseline in \cite{chen2018uniform} is twice as those of our method and the one in \cite{wang2020optimized}.

    \begin{figure}[t]
		\centering
		\includegraphics[width=0.95\linewidth]{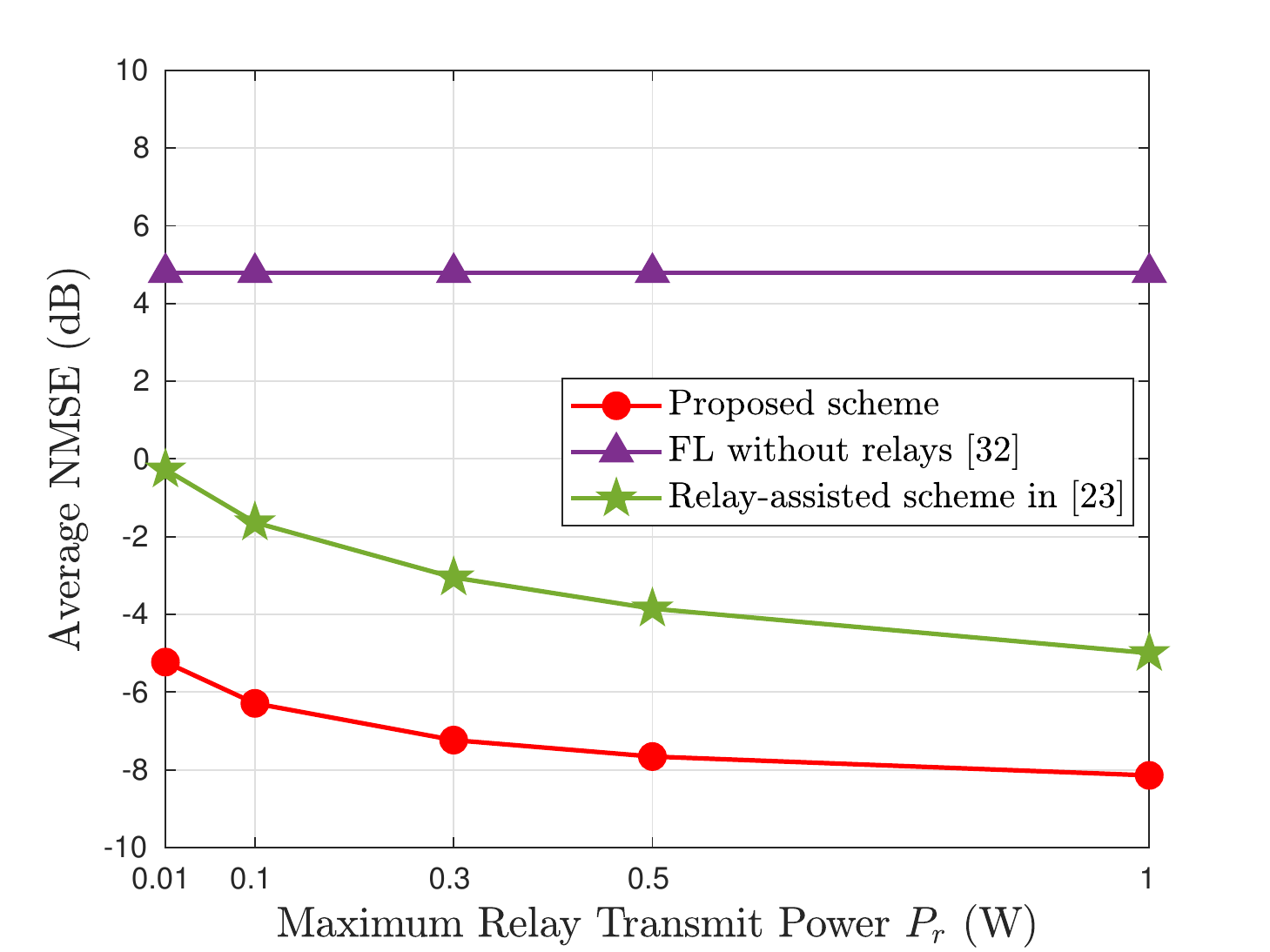}
		\caption{Average NMSE versus $P_r$ with $K = 20, N = 1$.} \label{fig:nmse_Pr_N1}
	\end{figure}
    \begin{figure}[t]
		\centering
		\includegraphics[width=0.95\linewidth]{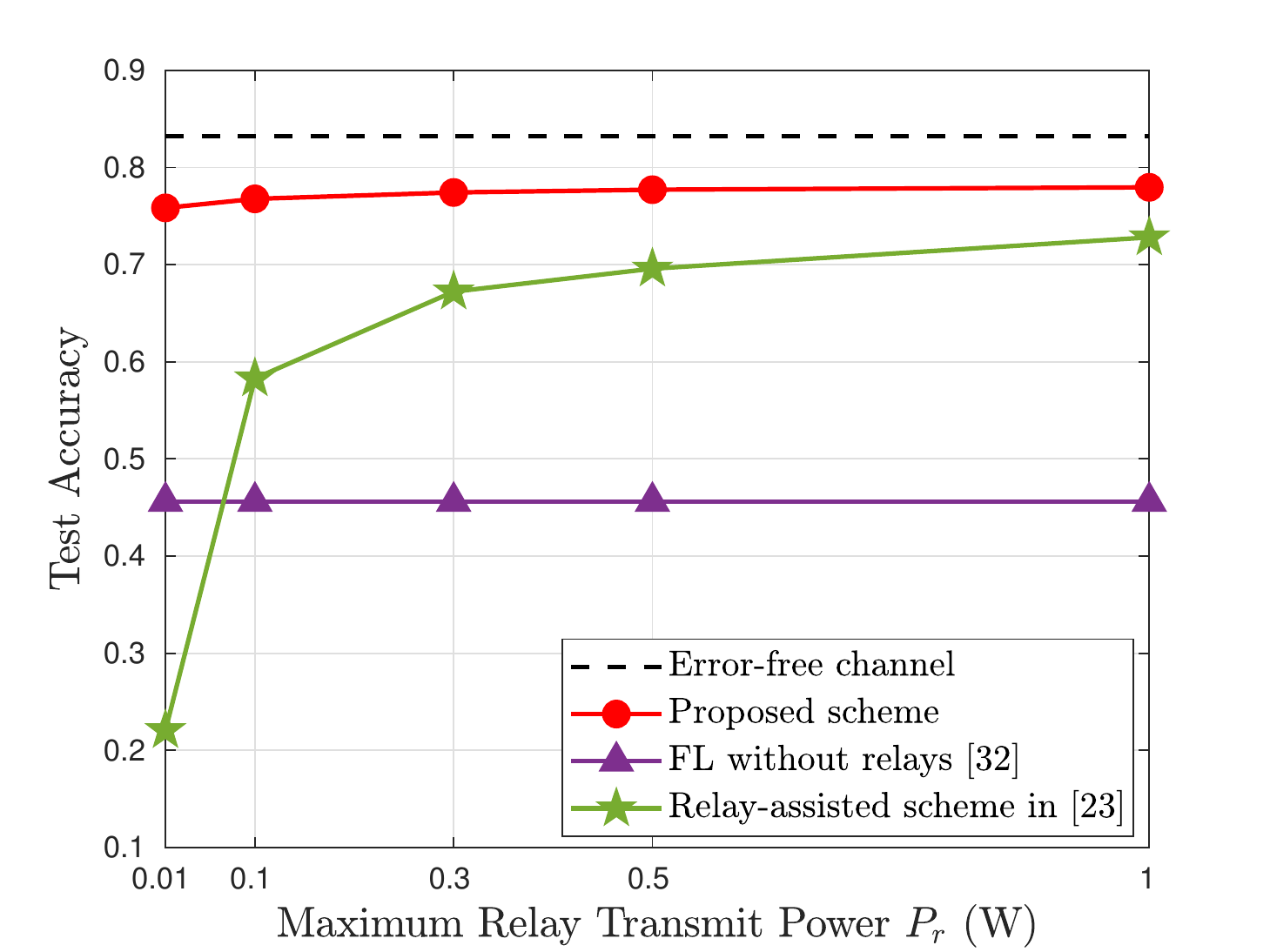}
		\caption{Test accuracy versus $P_r$ with $K = 20, N = 1$.} \label{fig:cmp_Pr_N1}
	\end{figure}

    \begin{figure}[t]
		\centering
		\includegraphics[width=0.95\linewidth]{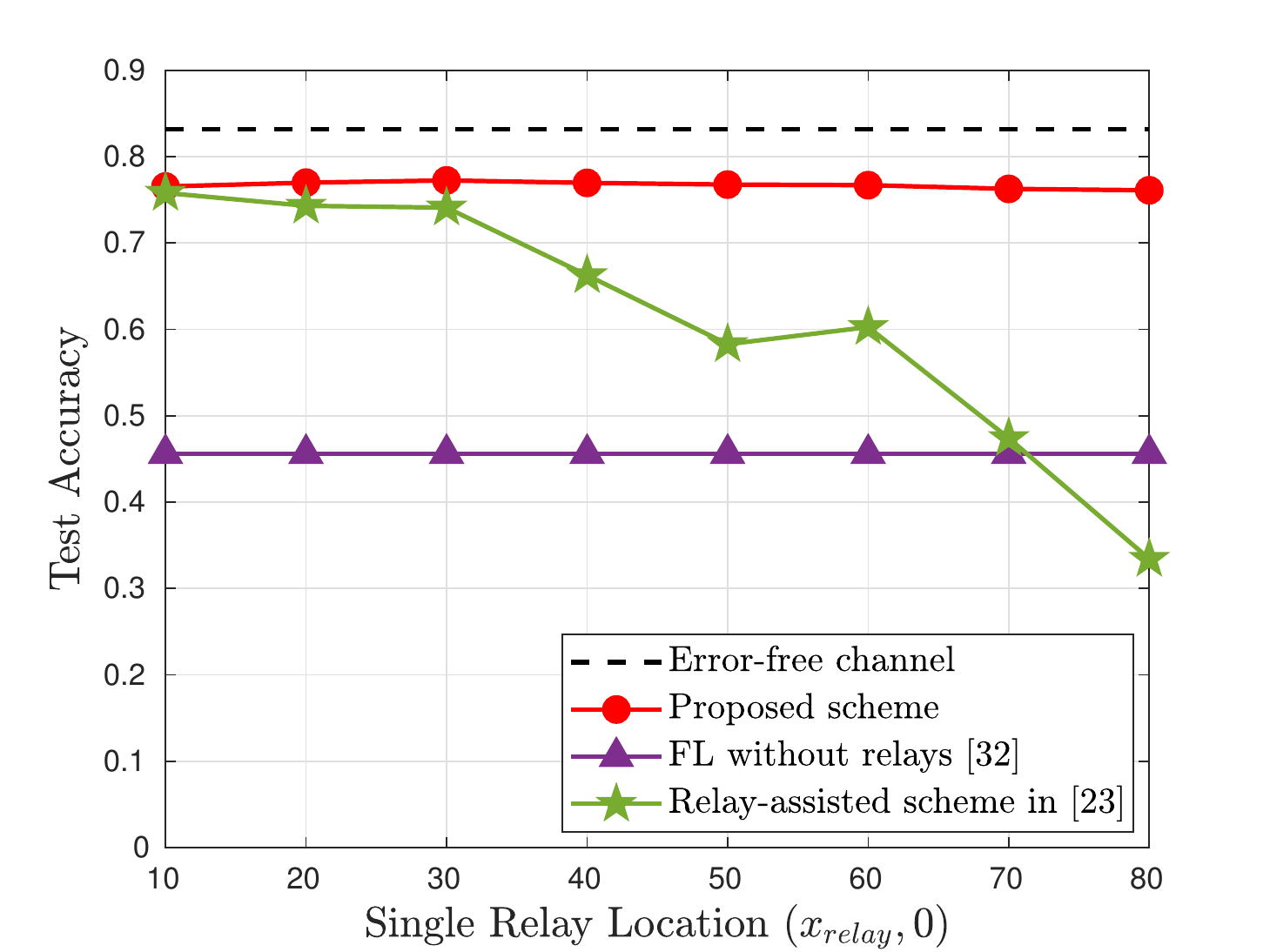}
		\caption{Test accuracy versus the single relay location with $K = 20, P_r = 0.1$ W, $P_0 = 0.05$ W.} \label{fig:cmp_loc_N1_Pr01}
	\end{figure}

    \begin{figure}[t]
		\centering
		\includegraphics[width=0.95\linewidth]{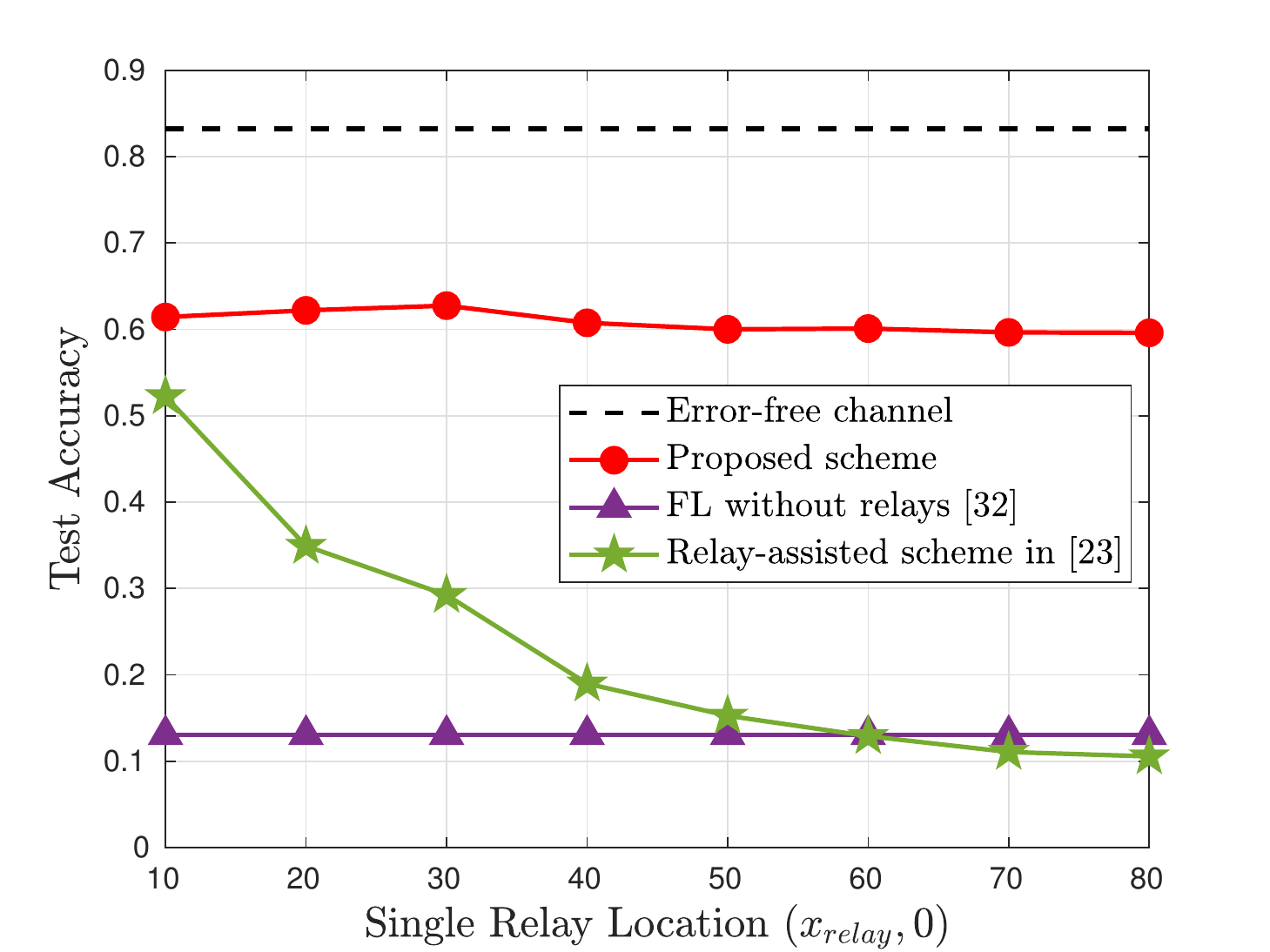}
		\caption{Test accuracy versus the single relay location with $K = 20, P_r = 0.01$ W,  $P_0 = 0.005$ W.} \label{fig:cmp_loc_N1_Pr001}
	\end{figure}
	\subsection{Simulations on Single Relay}
	In this subsection, we evaluate the FL performance of the proposed scheme in the single-relay case, i.e., $N = 1$. Specifically, we consider a two-dimensional plane where the AP and the single relay are placed at $(0, 0)$ and $(x_{\text{relay}}, 0)$ with $x_{\text{relay}} = 50$, respectively, and the $K$ devices are uniformly distributed within a rectangular region such that the location $(x_k, y_k)$ of each device $k$ satisfies $80 \leq x_k \leq 120$ and $-60 \leq y_k \leq 60$.
	
    We first evaluate the FL performance of different schemes in the case that the SNR is sufficiently high, where the noise power $\sigma^2$ is set to $-100$ dBm. The model aggregation error and test accuracy are presented in Table \ref{table:NMSE_SNR130} and Fig. \ref{fig:cmp_time_N1_SNR130}, respectively, where $1,000$ transmission blocks are used. We characterize the model aggregation error by the normalized MSE (NMSE), which is defined as
	\begin{equation}
		\text{NMSE} \triangleq \frac{\Big\|\widehat{\sum_{k = 1}^K \rho_k \bm{\Delta}_{k}} - \sum_{k = 1}^K \rho_k \bm{\Delta}_{k} \Big\|^2}{\big\|\sum_{k = 1}^K \rho_k \mathbf{\Delta}_k \big\|^2}.
	\end{equation}
	From Table \ref{table:NMSE_SNR130}, we see that all schemes can achieve an average NMSE smaller than $-20$ dB. As shown in Fig. \ref{fig:cmp_time_N1_SNR130}, such a small aggregation error has a negligible impact on the FL performance. On the other hand, the relay-assisted schemes, i.e., the proposed scheme and the baseline in \cite{wang2020optimized}, nearly double the number of transmission blocks to achieve the same test accuracy as the baseline without relays.
	
	We next evaluate the FL performance of different schemes with $\sigma^2$ equals to $-70$ dBm in Table \ref{table:NMSE} and Fig. \ref{fig:cmp_time_N1}. We see that the baseline without relays suffers a large aggregation error in this case. As the training proceeds, the accumulated aggregation error leads to an increasingly inaccurate global model in this scheme, which degrades the test accuracy. Thanks to the effect of relaying, the baseline in \cite{wang2020optimized} achieves a relatively smaller NMSE, but it still suffers from a low convergence rate. In contrast, our relay-assisted scheme has a much smaller NMSE and achieves significant accuracy improvement compared with these baselines.

	The above simulation results show that our relay-assisted scheme has significant performance gains in the low-to-intermediate SNR regime. By reducing the model aggregation error, the proposed scheme takes much less time to converge and achieves a much higher accuracy than the scheme without relays, even though it requires two transmission blocks in each aggregation  iteration.
	
    In Figs. \ref{fig:nmse_Pr_N1} and \ref{fig:cmp_Pr_N1}, we study the impact of the maximum relay transmit power $P_r$ on the model aggregation error and test accuracy with $1,000$ transmission blocks. We see that the proposed design outperforms the two baselines and achieves a near-optimal accuracy. The results of the baseline in \cite{chen2018uniform} are invariant to $P_r$ because it does not exploit the relay. The baseline in \cite{wang2020optimized} achieves a better NMSE performance than the baseline without relays \cite{chen2018uniform} when $P_r \geq 0.01$ W. However, when $P_r = 0.01$ W, the baseline in \cite{wang2020optimized} fails due to its strong reliance on the relay. In contrast, the test accuracy of the proposed design is insensitive to the variation of $P_r$ since it further exploits the performance gain from the direct transmissions from the devices to the AP in the two phases.

	In Figs. \ref{fig:cmp_loc_N1_Pr01} and \ref{fig:cmp_loc_N1_Pr001}, we investigate the impact of the relay location on the FL performance with high transmit power ($P_r = 0.1$ W, $P_0 = 0.05$ W) and low transmit power ($P_r = 0.01$ W, $P_0 = 0.005$ W), respectively. Specifically, we change the location of the single relay by configuring $x_{\text{relay}}$. The baseline in \cite{wang2020optimized} achieves significant performance improvement when the single relay is close to the AP, but the test accuracy tends to decrease as the single relay is farther away from the AP and finally becomes worse than the scheme without relays. In contrast, the effect of the relay location on the test accuracy of the proposed scheme is marginal and almost negligible. Therefore, we can conclude that the proposed scheme is robust to the relay locations.

	\begin{figure}[t]
		\centering
		\includegraphics[width=0.95\linewidth]{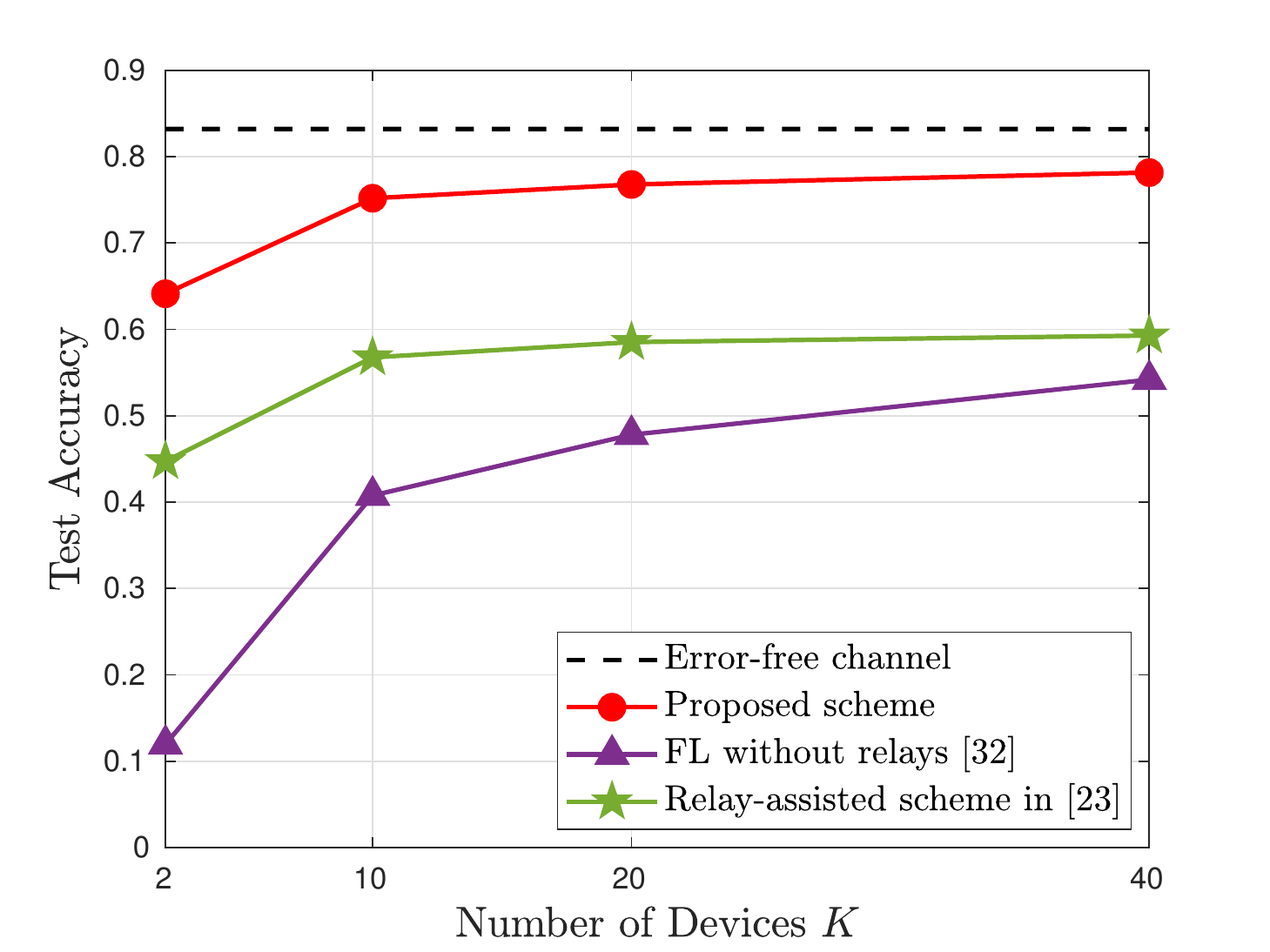}
		\caption{Test accuracy versus $K$ with $N = 1, P_r = 0.1$ W.} \label{fig:cmp_K_N1}
	\end{figure}
	
	In Fig. \ref{fig:cmp_K_N1}, we study the impact of the number of devices $K$ on the FL performance. Here, we fix the total number of training samples $D = 60000$ and assign $60000 / K$ samples to each device. Therefore, a larger $K$ means a smaller number of local training data and thus a smaller aggregation weight $\rho_k$ for each device. When $K$ is small, we observe dramatic accuracy improvements for all the algorithms as $K$ increases. This is because increasing the number of local devices decreases the weights of devices (i.e., $\{\rho_k\}$ in \eqref{global_model}), which mitigates the negative effect of potential deep fading on model aggregation. We see from Fig. \ref{fig:cmp_K_N1} that our algorithm outperforms the two baselines no matter what the value of $K$ is.
	
    \begin{figure}[t]
		\centering
		\includegraphics[scale=0.35]{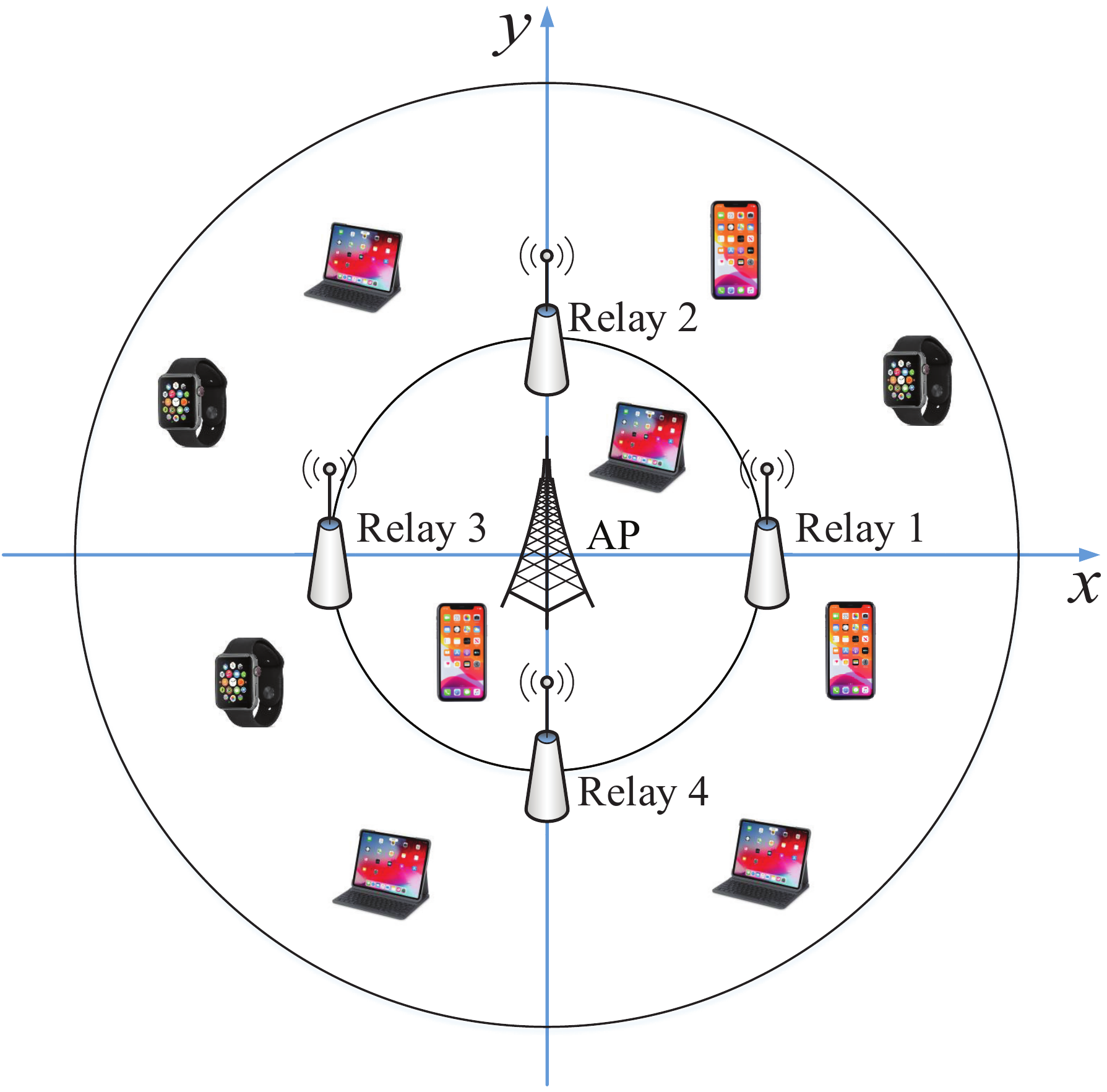}
		\caption{A single-cell relay-assisted FL system.} \label{fig:cell}
	\end{figure}
    \begin{figure}[t]
        \centering
		\includegraphics[width=0.95\linewidth]{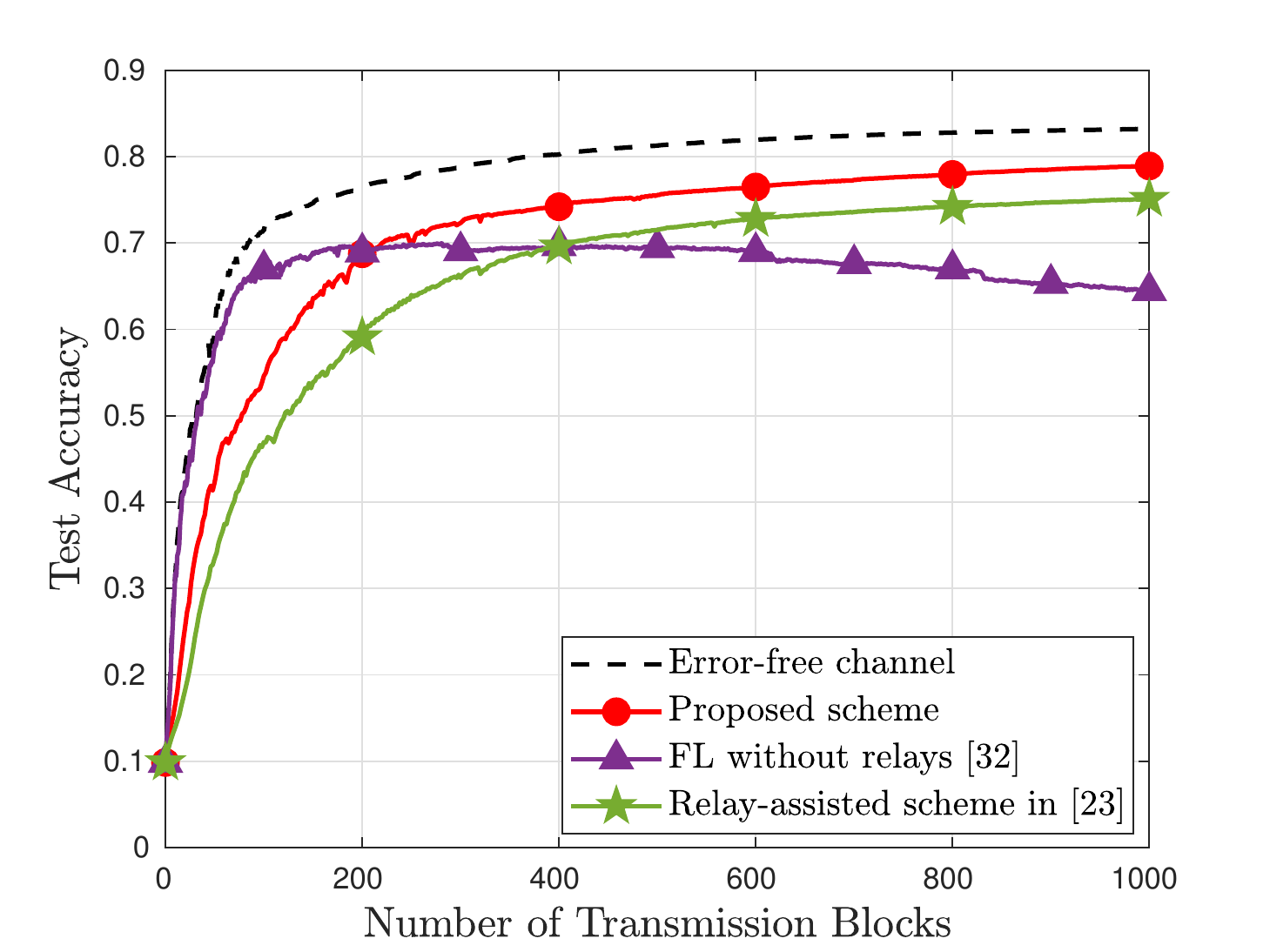}
		\caption{Test accuracy with $K = 20, N = 4, P_r = 0.1$ W.} \label{fig:cmp_time_N4}
	\end{figure}
	\subsection{Simulations on Multiple Relays}
	In this subsection, we evaluate the performance of the proposed scheme in the scenarios with multiple relays. In particular, we simulate a single-cell network with the radius of $120$ meters as shown in Fig. \ref{fig:cell}, where the AP is located at the center, the $K$ devices are uniformly distributed in the cell, and the $N$ relays are uniformly located at equal spacing on a circle centered at the AP with the radius of $50$ meters. We set $N = 4$ unless otherwise stated.
	
	\begin{figure}[t]
		\centering
		\includegraphics[width=0.95\linewidth]{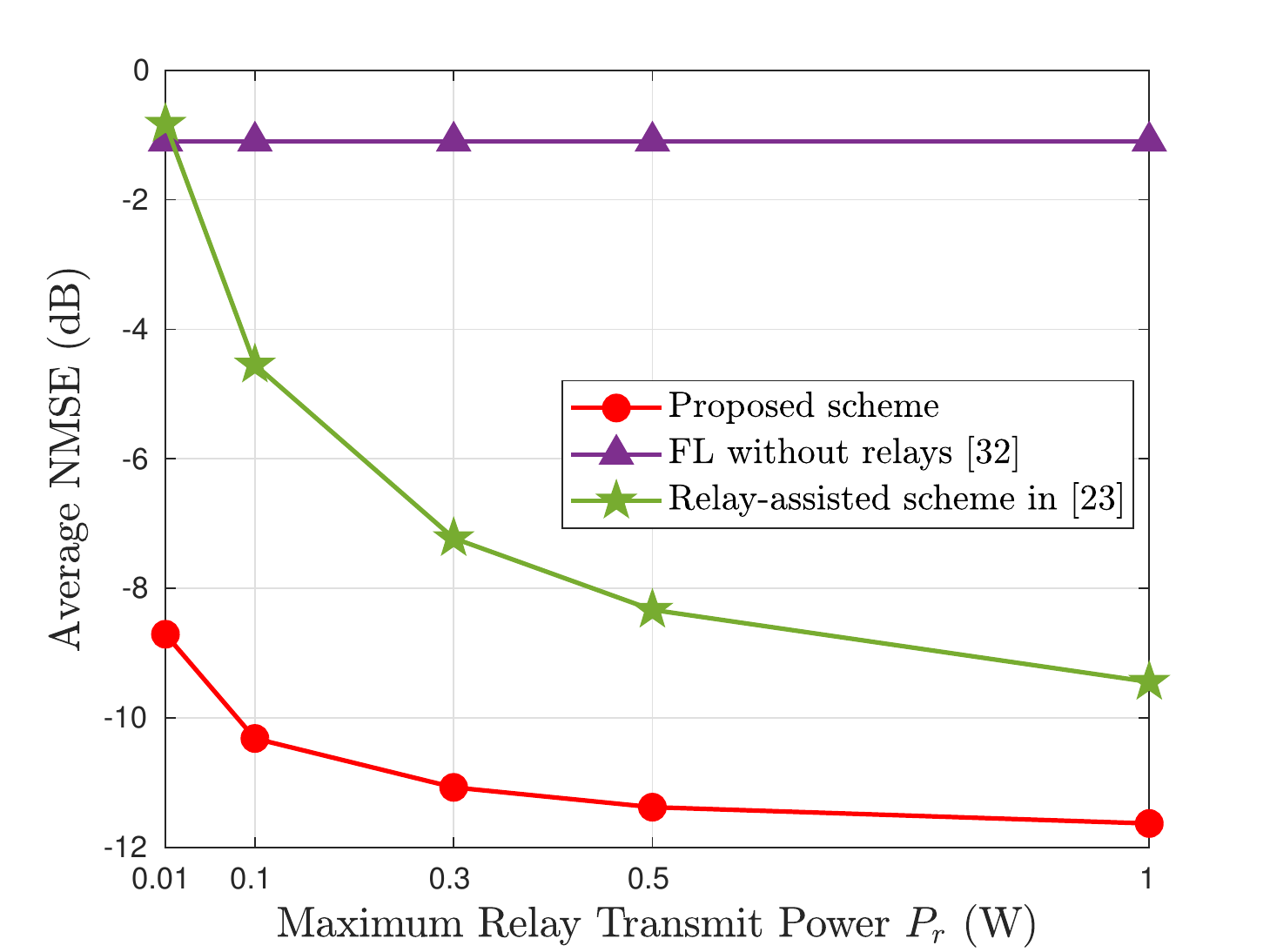}
		\caption{Average NMSE versus $P_r$ with $K = 20, N = 4$.} \label{fig:nmse_Pr_N4}
	\end{figure}
    \begin{figure}[t]
		\centering
		\includegraphics[width=0.95\linewidth]{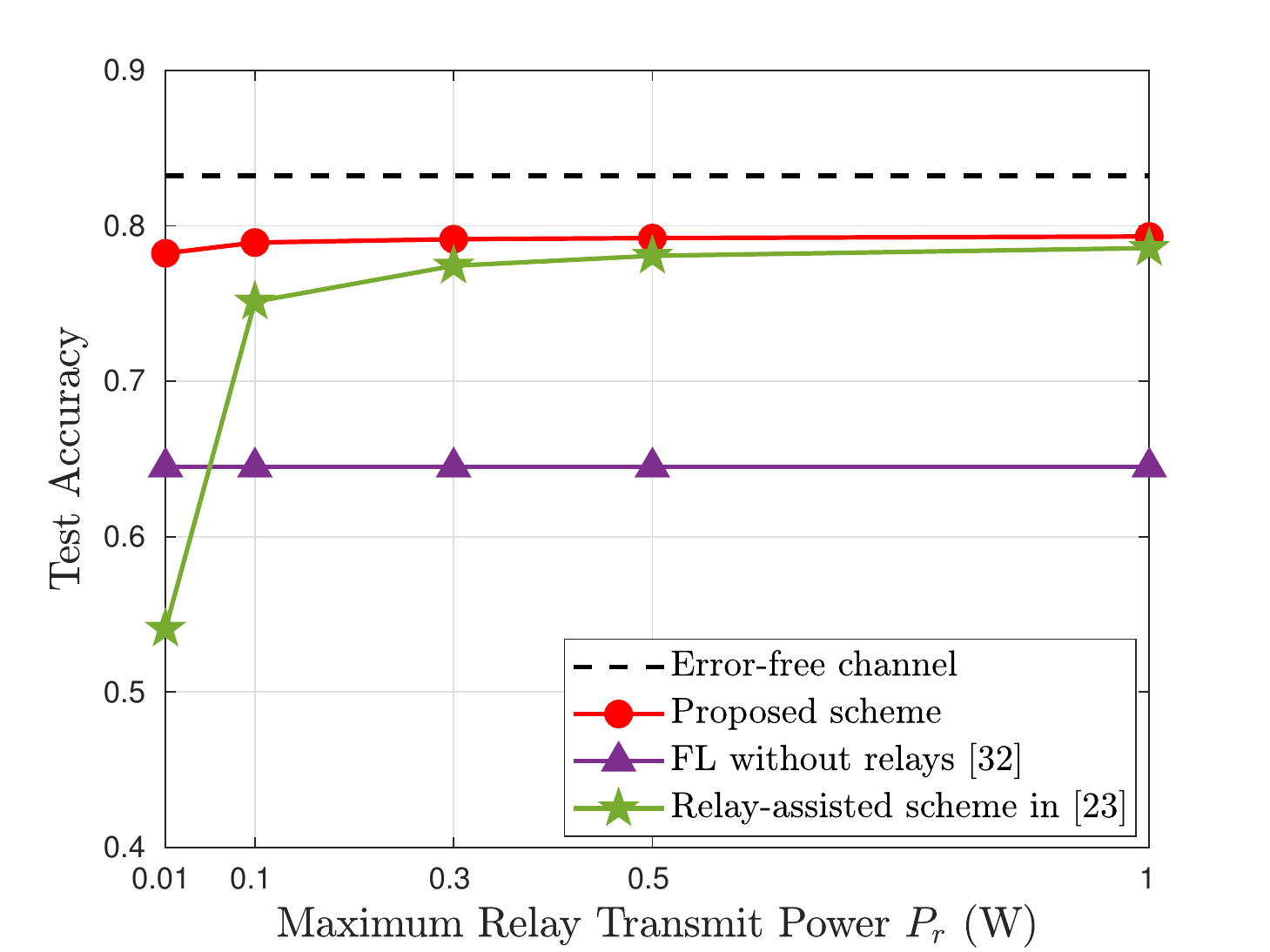}
		\caption{Test accuracy versus $P_r$ with $K = 20, N = 4$.} \label{fig:cmp_Pr_N4}
	\end{figure}

	In Fig. \ref{fig:cmp_time_N4}, we plot the test accuracy in $1,000$ transmission blocks. Similar to the single-relay case, the conventional scheme suffers from accuracy degradation because of its large model aggregation error. As shown by the red and green curves, the relays help overcome such an issue. Moreover, compared with the baseline in \cite{wang2020optimized}, our method further improves the learning accuracy by exploiting the cooperation of the devices and the relays.
	
    \begin{figure}[t]
		\centering
		\includegraphics[width=0.95\linewidth]{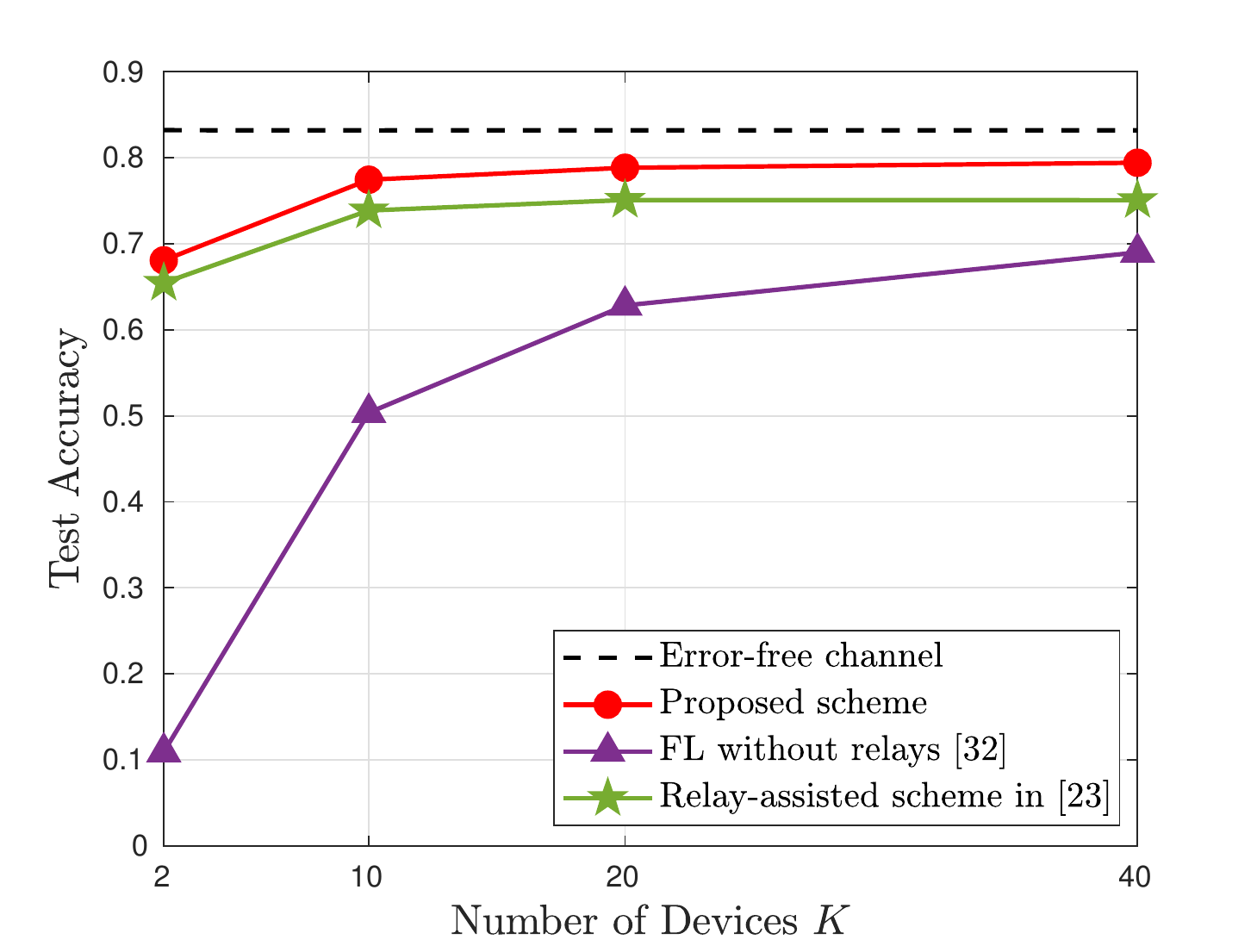}
		\caption{Test accuracy versus $K$ with $N = 4, P_r = 0.1$ W.} \label{fig:cmp_K_N4}
	\end{figure}
    \begin{figure}[t]
		\centering
		\includegraphics[width=0.95\linewidth]{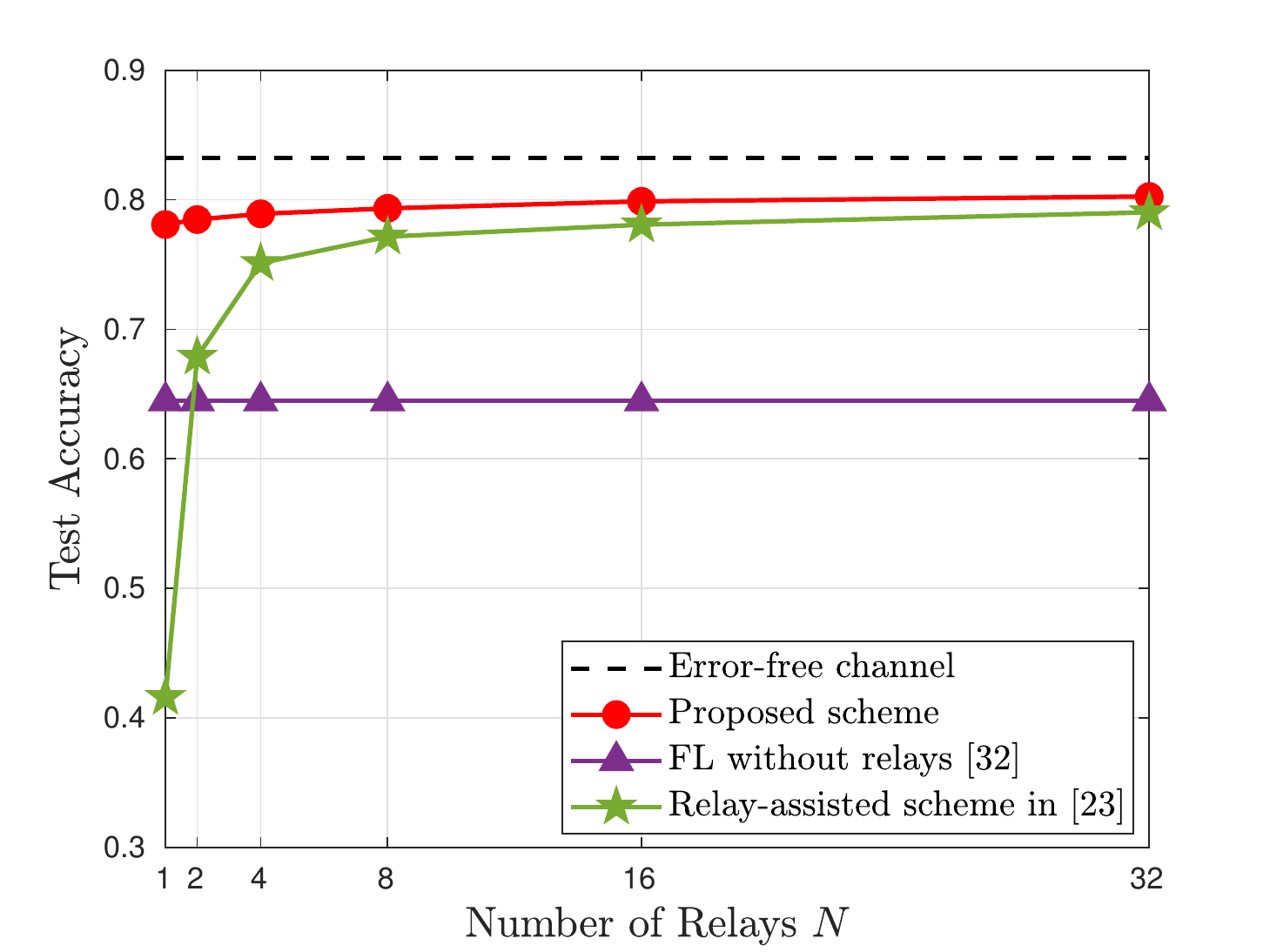}
		\caption{Test accuracy versus $N$ with $K = 20, P_r = 0.1$ W.} \label{fig:cmp_N}
	\end{figure}
	\begin{figure}[t]
		\centering
		\includegraphics[width=0.95\linewidth]{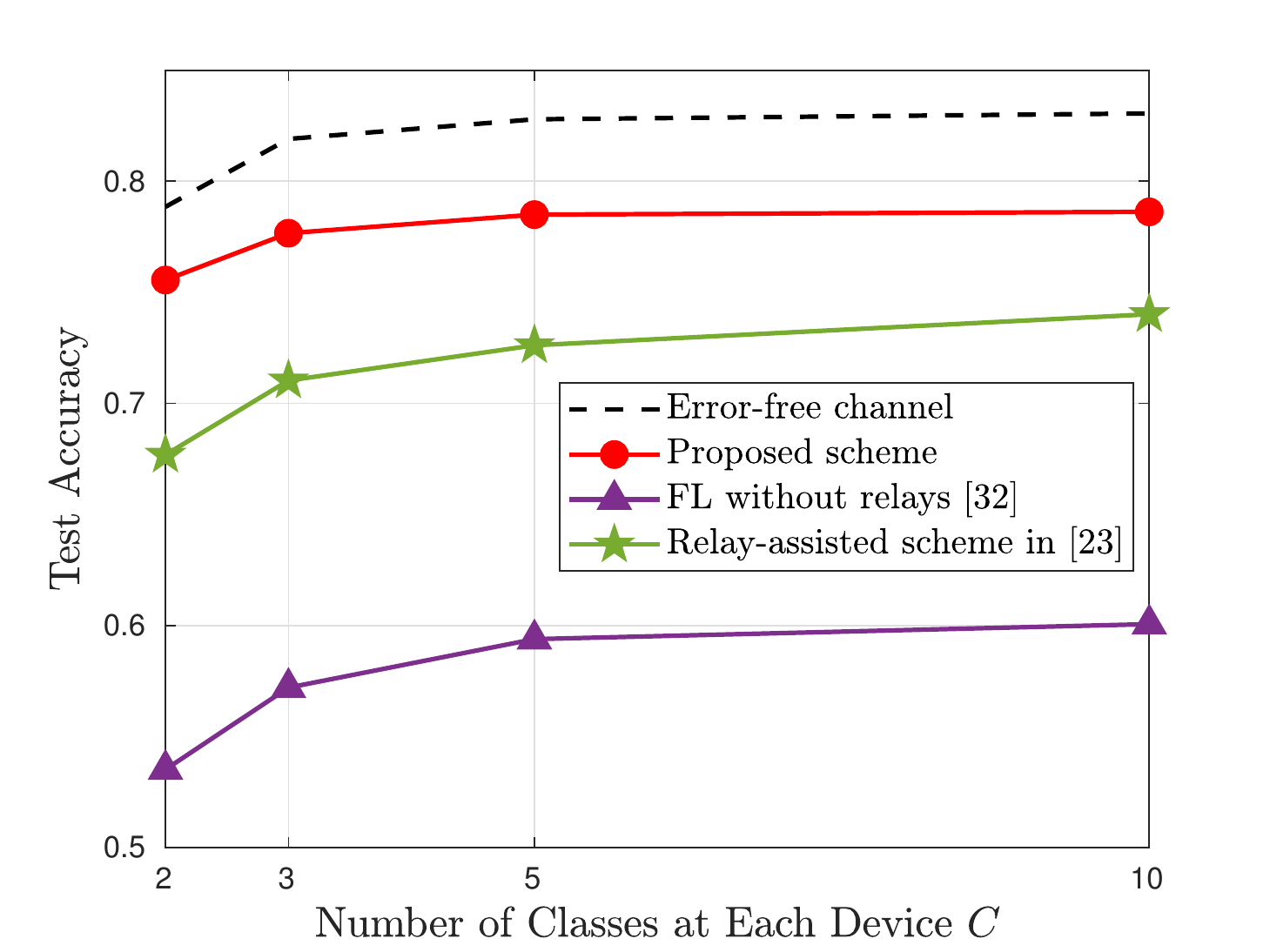}
		\caption{Test accuracy versus $C$ under non-i.i.d. data distribution with $K = 20, N = 4, P_r = 0.1$ W.} \label{fig:cmp_noniid_N4}
	\end{figure}
    \begin{figure}[t]
		\centering
		\includegraphics[width=0.95\linewidth]{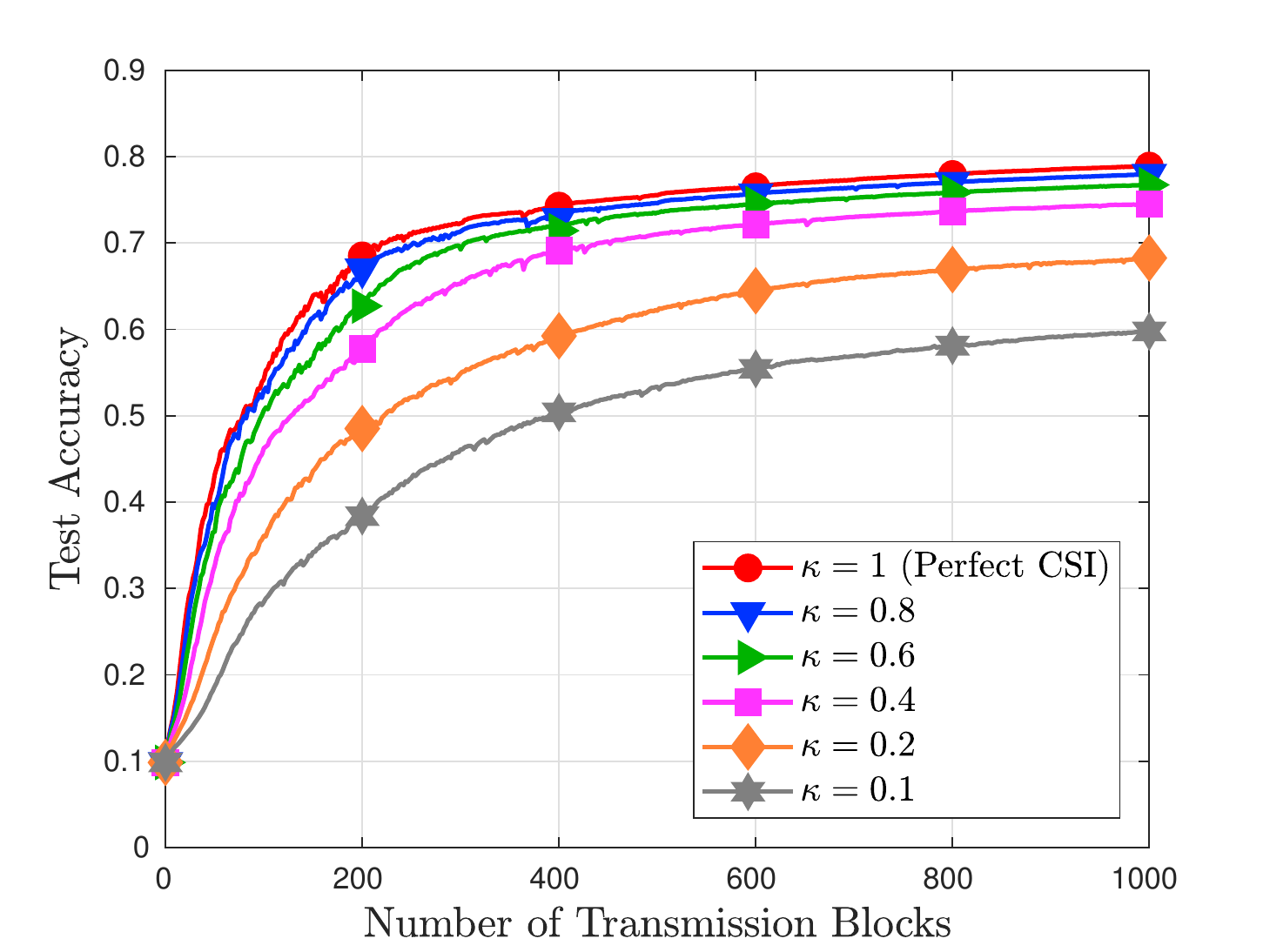}
		\caption{Test accuracy of the proposed scheme under different levels of CSI error with $K = 20, N = 4, P_r = 0.1$ W.
		} \label{fig:cmp_CSIE_N4}
	\end{figure}
	Figs. \ref{fig:nmse_Pr_N4} and \ref{fig:cmp_Pr_N4} plot the model aggregation error and test accuracy, respectively, under various choices of the maximum relay transmit power $P_r$ with $1,000$ transmission blocks. Similarly to the single-relay case, the proposed scheme outperforms the two baselines, and its corresponding test accuracy is insensitive to $P_r$. With the assistance of a number of relays, the NMSE of the baseline in \cite{wang2020optimized} significantly decreases with $P_r$, which leads to increasingly higher accuracy as shown in Fig. \ref{fig:cmp_Pr_N4}. When $P_r \geq 0.5$ W,  the baseline in \cite{wang2020optimized} performs similar to our method in terms of test accuracy owing to the availability of high relay power.
	
	In Figs. \ref{fig:cmp_K_N4} and	\ref{fig:cmp_N}, we plot the test accuracy versus the number of devices $K$ and the number of relays $N$, respectively. The baseline in \cite{wang2020optimized} performs close to the proposed design when $K$ is small. However, the performance gap becomes slightly larger as $K$ increases, verifying the additional performance gain from the relay-device cooperation in our design. In Fig. \ref{fig:cmp_N}, we see that the performance of the baseline in \cite{wang2020optimized} highly depends on the number of relays $N$, and this baseline fails with a small $N$ (i.e., $N \leq 4$). In contrast, the proposed design is less sensitive to the increase of $N$, verifying the robustness of the proposed design. Meanwhile, the results of the baseline in \cite{chen2018uniform} are invariant to the number of relays $N$ because it does not consider relays.

	\subsection{Simulations on Non-i.i.d. Data Distribution}
	Next, we consider non-i.i.d. data distribution and study the impact of data heterogeneity on the FL performance. We control the data heterogeneity by changing the number of classes of data at each device, denoted by $C$, with an equal number of training samples for each class. Specifically, we sort all training samples by their labels, divide them into $KC$ shards of size $\frac{60000}{KC}$, and assign each device with $C$ shards. As such, the training samples are distributed over the devices in a non-i.i.d. manner with the heterogeneity level represented by $C$. Specifically, a smaller $C$ means that the data distributions are more heterogeneous. As shown in Fig. \ref{fig:cmp_noniid_N4}, the value of $C$ critically affects the learning performance of all the FL schemes, especially when $C$ is small. Fig. \ref{fig:cmp_noniid_N4} also shows that the proposed relay-assisted scheme outperforms the baselines for heterogeneous data distributions.

	\subsection{Simulations on Imperfect CSI}
	Finally, we study the impact of CSI error on the FL performance. Specifically, we set the knowledge of the channel efficient as
	\begin{equation}
		\hat{h} = \sqrt{g} \left(\sqrt{\kappa} h + \sqrt{1 - \kappa} n \right)  \label{h_hat}
	\end{equation}
	where $h$ is the true channel coefficient, $n$ represents the CSI error following the standard i.i.d. Gaussian distribution, $g$ is the path loss, and $\kappa \in [0, 1]$ is a weighting factor that quantifies the CSI error. In \eqref{h_hat}, a smaller $\kappa$ means that the CSI error is larger. We optimize the transceiver and relay operation based on the knowledge of $\hat h$ given in \eqref{h_hat}. In Fig. \ref{fig:cmp_CSIE_N4}, we plot the test accuracy of the proposed scheme with different values of $\kappa$. When $\kappa \leq 0.4$, we observe evident accuracy degradation as the CSI error significantly enlarges the error in model aggregation. On the other hand, our scheme achieves comparable test accuracy with the case with perfect CSI when $\kappa \geq 0.6$. Fig. \ref{fig:cmp_CSIE_N4} shows that our proposed design is relatively robust against the CSI error if the CSI imperfection is limited.

	\section{Conclusions}
	In this paper, we proposed a novel relay-assisted cooperative model aggregation framework in FL systems and studied the transceiver design in this framework. We then formulated an optimization problem to jointly optimize the transmit scalars at the devices and relays and the de-noising receive scalars at the AP for model aggregation error minimization. We proposed an alternating minimization algorithm to solve the resulting non-convex problem with low complexity. Moreover, we rigorously analyzed the performance of the proposed relay-assisted cooperative FL design in a single-relay case and derived conditions on the relay such that our scheme achieves a smaller model aggregation error than the conventional scheme without relaying. The analysis provides a useful guideline on relay deployment to ensure beneficial relay assistance. Extensive simulations show that the proposed design achieves more accurate model aggregation and higher accuracy than the existing schemes. Furthermore, the proposed design is robust to the choices of the transmit power, the locations, and the number of relays as it exploits the cooperative diversity among the relays and devices.
	
	Finally, we conclude the paper with some interesting future directions of relay-assisted cooperative FL. First, the proposed cooperative FL scheme can be extended to systems with full-duplex relays, where the relays can simultaneously transmit and receive. In this case, we need to address new challenges, such as self-interference cancellation at the relays \cite{bharadia2013full} and additional model aggregation error caused by the delay in relaying the model signals. To this end, new designs on transceivers and relaying operation are needed to incorporate full-duplex relaying into cooperative FL. Second, it is challenging to extend the single-cell FL system to multi-cell communication systems as it requires further investigations on effective inter-cell interference mitigation as well as the relaying design \cite{yang2019scheduling}. Moreover, although the proposed relay-assisted design effectively reduces the model aggregation error, the proposed relaying operation inevitably amplifies and forwards the communication noise it has received. We shall investigate how to exploit de-noising techniques, such as relay-side decompression, to further improve the performance of relay-assisted FL.
	
	\appendices
	\section{Proof of Lemma \ref{lemmaA}} \label{appendixA}
	From \eqref{conventional_mse}, we have
	\begin{equation}
		\bar{e}_{\text{no-relay}} = \sum_{k = 1}^K \left|c h_{k} a_{k} - \rho_k \right|^2  + |c|^2 \sigma^2 \geq |c|^2 \sigma^2,  \label{MSE_A}
	\end{equation}
	where the equality holds when $c h_{k} a_{k} - \rho_k = 0, \forall k$. In other words, the minimum value of $\bar{e}_{\text{no-relay}}$ is achieved when the transmit scalar at each device is aligned with its corresponding model weight $\rho_k$. Therefore, the optimal $\{a_k\}$ is given by
	\begin{equation} \label{opt_a_A}
		a_{k} = \frac{\rho_k}{c h_k }, ~~\forall k \in \mathcal{K}.
	\end{equation}
	By \eqref{opt_a_A} and \eqref{conventional_cons}, we have
	\begin{equation}
		|a_k|^2 = \frac{\rho_k^2}{|c|^2 |h_k|^2 } \leq 2 P_0,  ~~\forall k \in \mathcal{K},
	\end{equation}
	which yields
	\begin{equation}
		|c|^2 \geq \frac{\rho_k^2}{2 P_0 |h_k|^2}, ~~\forall k \in \mathcal{K} \Leftrightarrow |c| \geq \frac{1}{\sqrt{2 P_0}} \max_{k \in \mathcal{K}} \frac{\rho_k}{|h_k|}.  \label{c_A}
	\end{equation}
	From \eqref{MSE_A}, we have
	\begin{equation}
		\bar{e}_{\text{no-relay}} = |c|^2 \sigma^2,  \label{bar_e}
	\end{equation}
	when $c h_{k} a_{k} - \rho_k = 0, \forall k$. Therefore, $\bar{e}_{\text{no-relay}}$ increases with $|c|$ and the minimum $\bar{e}_{\text{no-relay}}$ can be obtained by taking the minimum value of $|c|$. From \eqref{c_A}, we find that the minimum value of $|c|$ is
	\begin{equation}
		|c| = \frac{1}{\sqrt{2 P_0}} \max_{k \in \mathcal{K}} \frac{\rho_k}{|h_k|}.   \label{opt_c_A}
	\end{equation}
	Plugging \eqref{opt_c_A} into \eqref{bar_e}, we obtain the minimum $\bar{e}_{\text{no-relay}}$ as $\frac{\sigma^2}{2P_0} \max_{k \in \mathcal{K}} \frac{\rho_k^2}{|h_k|^2}$.

	\section{Proof of Theorem \ref{theorem1}} \label{appendixC}
	Here, we analyze a sub-optimal solution to \eqref{problem_single} by restricting $c_1 = 0$. With $c_1 = 0$, the remaining variables to be optimized are $\{a_{k, 1}, a_{k, 2}, b, c_2\}$. To facilitate our analysis, we introduce two auxiliary variables $\alpha \in \mathbb{R}$ and $\beta \in \mathbb{R}$ with $\alpha + \beta = 1$, and reformulate Problem \eqref{problem_single} as
	\begin{subequations} \label{problem_single2}
		\begin{align}
			\min_{\{a_{k, 1}, a_{k, 2}, b, c_2, \alpha, \beta\}}&\sum_{k = 1}^K \left| c_2 f b g_{k} a_{k, 1} + c_2 h_{k} a_{k, 2} - \left(\alpha + \beta \right) \rho \right|^2 \nonumber\\
            &+ |c_2|^2 \left(1 + |f|^2 |b|^2 \right) \sigma^2  \label{obj_single2} \\
			{\rm s.t.}~~~~~~
			&\eqref{cons1_1}, \eqref{cons1_2}, \eqref{cons_single},  \nonumber \\
			&\alpha + \beta = 1. \label{cons_single2}
		\end{align}
	\end{subequations}
	Problem \eqref{problem_single2} is still non-convex because all variables $\{a_{k, 1},$ $ a_{k, 2}, b, c_2, \alpha, \beta\}$ are coupled in the objective function \eqref{obj_single2} and constraint \eqref{cons_single} is non-convex. Notice that the objective of \eqref{problem_single2} satisfies that
	\begin{align}
		&\sum_{k = 1}^K \left| c_2 f b g_{k} a_{k, 1} + c_2 h_{k} a_{k, 2} - \left(\alpha + \beta \right) \rho \right|^2 \nonumber\\
            &+ |c_2|^2 \left(1 + |f|^2 |b|^2 \right) \sigma^2 \geq |c_2|^2 \left(1 + |f|^2 |b|^2 \right) \sigma^2,  \label{obj_cons}
	\end{align}
	where the equality holds when $c_2 f b g_{k} a_{k, 1} + c_2 h_{k} a_{k, 2} - \left(\alpha + \beta \right) \rho = 0, \forall k \in \mathcal{K}$. From \eqref{obj_cons}, we find that a sufficient condition to achieve the minimum objective in \eqref{obj_single2} is given by
	\begin{equation}  \label{a_single2}
		a_{k, 1} = \frac{\alpha \rho}{c_2 f b g_k}, \quad
		a_{k, 2} = \frac{\beta \rho}{c_2 h_k},  ~~\forall k \in \mathcal{K}.
	\end{equation}
	With \eqref{a_single2}, the original problem is simplified as
	\begin{subequations} \label{problem_single3}
		\begin{align}
			\min_{\{b, c_2, \alpha, \beta\}}~&|c_2|^2 \left(1 + |f|^2 |b|^2 \right) \sigma^2  \label{obj_single3} \\
			{\rm s.t.}~~~
			&\frac{\alpha^2 \rho^2}{|c_2|^2 |f|^2 |b|^2 |g_k|^2} \leq P_0,  ~~\forall k \in \mathcal{K},  \label{cons_single3_1}  \\
			&\frac{\beta^2 \rho^2}{|c_2|^2 |h_k|^2} \leq P_0,  ~~\forall k \in \mathcal{K},  \label{cons_single3_2}  \\
			&\frac{K \alpha^2 \rho^2}{|c_2|^2 |f|^2} + |b|^2 \sigma^2 \leq P_r,  \label{cons_single3_3}  \\
			&\alpha + \beta = 1. \label{cons_single3_4}
		\end{align}
	\end{subequations}
	Let $\eta = |c_2|^2$ and $\gamma = |c_2|^2 |b|^2$ with $|b|^2 = \gamma / \eta$. Substituting $\eta$ and $\gamma$ into \eqref{problem_single3}, we have
	\begin{subequations} \label{problem_single4}
		\begin{align}
			\min_{\{\eta, \gamma, \alpha, \beta\}}&\left(\eta + \gamma |f|^2 \right) \sigma^2  \label{obj_single4} \\
			{\rm s.t.}~~
			&\gamma \geq \frac{\alpha^2 \rho^2}{ P_0 |f|^2 |g_k|^2},  ~~\forall k \in \mathcal{K},  \label{cons_single4_1}  \\
			&\eta \geq \frac{\beta^2 \rho^2}{P_0 |h_k|^2},  ~~\forall k \in \mathcal{K},  \label{cons_single4_2}  \\
			&\eta \geq \frac{K \alpha^2 \rho^2 + \gamma \sigma^2 |f|^2}{P_r |f|^2},  \label{cons_single4_3}  \\
			&\alpha + \beta = 1. \label{cons_single4_4}
		\end{align}	
	\end{subequations}
	
	Note that Problem \eqref{problem_single4} jointly convex in $(\eta, \gamma)$ once $\alpha$ and $\beta$ are given. Thus, we express the optimal $\gamma^*$ and $\eta^*$ in \eqref{problem_single4} as the functions of the auxiliary variables $\alpha, \beta$, which are given by
		\begin{align}
			\gamma^* &= \frac{\alpha^2 \rho^2}{ P_0 |f|^2 } \max_{k \in \mathcal{K}} \frac{1}{|g_k|^2}, \\
			\eta^* &= \max \left\{ \frac{\beta^2 \rho^2}{P_0} \max_{k \in \mathcal{K}} \frac{1}{|h_k|^2}, \frac{K \alpha^2 \rho^2 + \gamma^* \sigma^2 |f|^2}{P_r |f|^2}  \right\}.  \label{opt_eta}
		\end{align}
	We here consider two cases:
	\begin{enumerate}
		\item Case 1: If $\frac{\beta^2 \rho^2}{P_0} \max_{k \in \mathcal{K}} \frac{1}{|h_k|^2} \geq \frac{K \alpha^2 \rho^2 + \gamma^* \sigma^2 |f|^2}{P_r |f|^2},$ we have $\eta^* = \frac{\beta^2 \rho^2}{P_0} \max_{k \in \mathcal{K}} \frac{1}{|h_k|^2}$. By noting $\alpha + \beta = 1$ and $\gamma^* = \frac{\alpha^2 \rho^2}{ P_0 |f|^2 \min_{k \in \mathcal{K}} |g_k|^2}$, we equivalently rewrite the above condition as
		\begin{equation}
			\frac{(1 - \alpha)^2 \rho^2}{P_0 \min\limits_{k \in \mathcal{K}} |h_k|^2} \geq \frac{K \alpha^2 \rho^2 P_0 \min\limits_{k \in \mathcal{K}} |g_k|^2 + \sigma^2 \alpha^2 \rho^2}{P_r |f|^2 P_0 \min\limits_{k \in \mathcal{K}} |g_k|^2},  \label{quad_ineq1}
		\end{equation}
		which is equivalent to
		\begin{align}
			0 \leq& \alpha^2 \Big(P_r |f|^2 \min_{k \in \mathcal{K}} |g_k|^2 - K P_0 \min_{k \in \mathcal{K}} |g_k|^2 \min_{k \in \mathcal{K}} |h_k|^2 \nonumber\\
            &- \sigma^2 \min_{k \in \mathcal{K}} |h_k|^2 \Big) - 2 \alpha P_r |f|^2 \min_{k \in \mathcal{K}} |g_k|^2 \nonumber\\
            &+ P_r |f|^2 \min_{k \in \mathcal{K}} |g_k|^2.  \label{quad_ineq2}
		\end{align}
		As $0 \leq \alpha \leq 1$, we can further express \eqref{quad_ineq2} as
		\begin{equation}
			0 \leq \alpha \leq \frac{1}{1 + \sqrt{\frac{ \left(K P_0 \min_{k \in \mathcal{K}} |g_k|^2 + \sigma^2 \right) \min_{k \in \mathcal{K}} |h_k|^2}{P_r |f|^2 \min_{k \in \mathcal{K}} |g_k|^2}}}. \label{cond_alpha1}
		\end{equation}
		\item Case 2: If $\frac{\beta^2 \rho^2}{P_0} \max_{k \in \mathcal{K}} \frac{1}{|h_k|^2} \leq \frac{K \alpha^2 \rho^2 + \gamma^* \sigma^2 |f|^2}{P_r |f|^2},$ we have $\eta^* = \frac{K \alpha^2 \rho^2 + \gamma^* \sigma^2 |f|^2}{P_r |f|^2}$. Similarly, we can equivalently rewrite the condition of this case as
		\begin{equation}
			\frac{(1 - \alpha)^2 \rho^2}{P_0 \min\limits_{k \in \mathcal{K}} |h_k|^2} \leq \frac{K \alpha^2 \rho^2 P_0 \min\limits_{k \in \mathcal{K}} |g_k|^2 + \sigma^2 \alpha^2 \rho^2}{P_r |f|^2 P_0 \min\limits_{k \in \mathcal{K}} |g_k|^2},  \label{quad_ineq3}
		\end{equation}
		which is equivalent to
		\begin{equation}
			\frac{1}{1 + \sqrt{\frac{ \left(K P_0 \min_{k \in \mathcal{K}} |g_k|^2 + \sigma^2 \right) \min_{k \in \mathcal{K}} |h_k|^2}{P_r |f|^2 \min_{k \in \mathcal{K}} |g_k|^2}}} \leq \alpha \leq 1.  \label{cond_alpha2}
		\end{equation}
	\end{enumerate}
	Define $\bar{\alpha} \triangleq \frac{1}{1 + \sqrt{\frac{ \left(K P_0 \min_{k \in \mathcal{K}} |g_k|^2 + \sigma^2 \right) \min_{k \in \mathcal{K}} |h_k|^2}{P_r |f|^2 \min_{k \in \mathcal{K}} |g_k|^2}}}$. We summarize the above two cases in \eqref{cond_alpha1} and \eqref{cond_alpha2} to express the optimal $\eta^*$ as
	\begin{align}
		\eta^* =
		\begin{cases}
			\frac{\beta^2 \rho^2}{P_0} \max_{k \in \mathcal{K}} \frac{1}{|h_k|^2},  &\mbox{if $0 \leq \alpha \leq \bar{\alpha}$},  \\
			\frac{K \alpha^2 \rho^2 + \gamma^* \sigma^2 |f|^2}{P_r |f|^2},  &\mbox{if $\bar{\alpha} \leq \alpha \leq 1$}.
		\end{cases}
	\end{align}
	Therefore, the value in \eqref{obj_single4} is given by
	\begin{align}
		&\min \left(\eta + \gamma |f|^2 \right) \sigma^2 = \nonumber\\
        &
		\begin{cases}
			\frac{\alpha^2 \rho^2 \sigma^2}{ P_0 } \max\limits_{k \in \mathcal{K}} \frac{1}{|g_k|^2} + \frac{\beta^2 \rho^2 \sigma^2 }{P_0} \max\limits_{k \in \mathcal{K}} \frac{1}{|h_k|^2},  &\mbox{if $0 \leq \alpha \leq \bar{\alpha}$},  \\
			\frac{\alpha^2 \rho^2 \sigma^2}{ P_0 } \max\limits_{k \in \mathcal{K}} \frac{1}{|g_k|^2} \bigg(1 + \frac{K P_0 \min\limits_{k \in \mathcal{K}} |g_k|^2 + \sigma^2 }{P_r |f|^2} \bigg),  &\mbox{if $\bar{\alpha} \leq \alpha \leq 1$}.
		\end{cases}
	\end{align}
	
	On one hand, when $0 \leq \alpha \leq \bar{\alpha}$, we have
	\begin{align}
		\left(\eta + \gamma |f|^2 \right) \sigma^2 &= \frac{\alpha^2 \rho^2 \sigma^2}{ P_0 } \max\limits_{k \in \mathcal{K}} \frac{1}{|g_k|^2} + \frac{\beta^2 \rho^2 \sigma^2 }{P_0} \max\limits_{k \in \mathcal{K}} \frac{1}{|h_k|^2}  \label{case1} \\
		&\overset{\text{(iii)}}\geq \frac{\alpha^2 \rho^2 \sigma^2}{ P_0 } \max\limits_{k \in \mathcal{K}} \frac{1}{|g_k|^2} + \frac{\beta^2 \rho^2 \sigma^2 }{P_0} \max\limits_{k \in \mathcal{K}} \frac{1}{|g_k|^2}  \nonumber \\
		&\overset{\text{(iv)}}\geq \frac{\rho^2 \sigma^2}{2 P_0 } \max\limits_{k \in \mathcal{K}} \frac{1}{|g_k|^2},
	\end{align}
	where (iii) is because we have $\min_{k \in \mathcal{K}} |g_k|^2 \geq \min_{k \in \mathcal{K}} |h_k|^2$ by condition \eqref{snr_relay}, and the equality in (iv) holds when $\alpha = \beta = \frac{1}{2}$. Substituting $\alpha = \beta = \frac{1}{2}$ into \eqref{case1}, we have
	\begin{align} \label{Pr_t11}
		\left(\eta + \gamma |f|^2 \right) \sigma^2 &= \frac{\rho^2 \sigma^2}{4 P_0 } \max\limits_{k \in \mathcal{K}} \frac{1}{|g_k|^2} + \frac{\rho^2 \sigma^2 }{4 P_0} \max\limits_{k \in \mathcal{K}} \frac{1}{|h_k|^2} \nonumber\\
    &\leq \frac{\rho^2 \sigma^2}{2 P_0 } \max\limits_{k \in \mathcal{K}} \frac{1}{|h_k|^2} = \bar{e}_{\text{no-relay}}.
	\end{align}
	Thus, it suffices to ensure that $\alpha = \beta = \frac{1}{2}$ can be satisfied when $0 \leq \alpha \leq \bar{\alpha}$, which gives
	\begin{align} \label{Pr_t12}
		\bar{\alpha} \geq \frac{1}{2} &\Leftrightarrow \frac{\left(K P_0 \min_{k \in \mathcal{K}} |g_k|^2 + \sigma^2 \right) \min_{k \in \mathcal{K}} |h_k|^2}{ P_r |f|^2 \min_{k \in \mathcal{K}} |g_k|^2} \leq 1  \nonumber \\
		&\Leftrightarrow \frac{P_r |f|^2}{\sigma^2} \geq K \min_{k \in \mathcal{K}} \frac{P_0 |h_k|^2}{\sigma^2} + \frac{ \min_{k \in \mathcal{K}} |h_k|^2}{ \min_{k \in \mathcal{K}} |g_k|^2}  \nonumber \\
		&\Leftrightarrow \text{SNR}_{\text{relay-AP}} \geq K \min\nolimits_{k \in \mathcal{K}} \text{SNR}_{k} + \delta.
	\end{align}
		
	On the other hand, when $\bar{\alpha} \leq \alpha \leq 1$, we need to make sure that
	\begin{align}  \label{Pr_t21}
		&\left(\eta + \gamma |f|^2 \right) \sigma^2 \nonumber \\
        =& \frac{\alpha^2 \rho^2 \sigma^2}{ P_0 } \max_{k \in \mathcal{K}} \frac{1}{|g_k|^2} \left(1 + \frac{K P_0 \min_{k \in \mathcal{K}} |g_k|^2 + \sigma^2 }{P_r |f|^2} \right) \nonumber \\
        \leq& \frac{\rho^2 \sigma^2}{2 P_0 } \max\limits_{k \in \mathcal{K}} \frac{1}{|h_k|^2} = \bar{e}_{\text{no-relay}},
	\end{align}
	which is equivalent to
	\begin{align} \label{Pr_t22}
		&\frac{P_r |f|^2}{\sigma^2} \geq \left( K \min\limits_{k \in \mathcal{K}} \frac{P_0 |h_k|^2}{\sigma^2} + \frac{ \min\limits_{k \in \mathcal{K}} |h_k|^2}{ \min\limits_{k \in \mathcal{K}} |g_k|^2} \right) \frac{2 \alpha^2}{1 - 2 \alpha^2 \frac{ \min\limits_{k \in \mathcal{K}} |h_k|^2}{ \min\limits_{k \in \mathcal{K}} |g_k|^2}}  \nonumber \\
		&\Leftrightarrow~
		\text{SNR}_{\text{relay-AP}} \geq \left(K \min\nolimits_{k \in \mathcal{K}} \text{SNR}_{k} + \delta \right) \frac{2 \alpha^2}{1 - 2 \alpha^2 \delta }.
	\end{align}
	Note that both the left hand side of the inequality in \eqref{Pr_t21} and the right hand side of \eqref{Pr_t22} increase with $\alpha$. Thus, the optimal $\alpha^*$ is given by $\alpha^* = \bar{\alpha}$ and we have
	\begin{equation} \label{Pr_t23}
		\text{SNR}_{\text{relay-AP}} \geq \left(K \min\nolimits_{k \in \mathcal{K}} \text{SNR}_{k} + \delta \right) \frac{2 \bar{\alpha}^2}{1 - 2 \bar{\alpha}^2 \delta }.
	\end{equation}
	
	By \eqref{Pr_t11} and \eqref{Pr_t21}, we find that $\left(\eta + \gamma |f|^2 \right) \sigma^2 \leq \bar{e}_{\text{no-relay}}$ can be achieved when at least one of the conditions \eqref{Pr_t12} and \eqref{Pr_t23} is satisfied, which leads to
	\begin{equation} \label{Pr_t24}
		\text{SNR}_{\text{relay-AP}} \geq \left(K \min_{k \in \mathcal{K}} \text{SNR}_{k} + \delta \right) \min \left\{1, \frac{2 \bar{\alpha}^2}{1 - 2 \bar{\alpha}^2 \delta } \right\}.
	\end{equation}
	Note that $\frac{2 \bar{\alpha}^2}{1 - 2 \bar{\alpha}^2 \delta } = \frac{2}{\left(1 + \sqrt{\frac{K \min_{k \in \mathcal{K}} \text{SNR}_k + \delta}{\text{SNR}_{\text{relay-AP}}}} \right)^2 - 2 \delta}$. We can equivalently express \eqref{Pr_t24} as
	\begin{equation}  \label{snr_tmp}
		\frac{1}{\varphi^2} \geq \min \left\{ 1, \frac{2}{(1 + \varphi)^2 - 2 \delta} \right\},
	\end{equation}
	where $\varphi \triangleq \sqrt{\frac{K \min_{k \in \mathcal{K}} \text{SNR}_k + \delta}{\text{SNR}_{\text{relay-AP}}}} \geq 0$.
	As $\delta \in [0, 1]$ when \eqref{snr_relay} holds, we can derive from \eqref{snr_tmp} that
	\begin{align}
		0 \leq \varphi \leq 1 + \sqrt{2 - 2 \delta}  ~\Rightarrow~ \frac{1}{\varphi^2} \geq \frac{1}{( 1 + \sqrt{2 - 2 \delta}  )^2},
	\end{align}
	which leads to condition \eqref{snr_final}.

	\bibliography{references}
\end{document}